\documentclass[prl,twocolumn,showpacs,amsmath,amssymb,superscriptaddress]{revtex4-1}
\usepackage{graphicx}
\usepackage{epsfig,epstopdf} 
\usepackage{dcolumn}
\usepackage{bm}
\usepackage{bbm}
\usepackage{ulem}
\usepackage{color}
\usepackage{slashed}
\usepackage{hyperref}
\usepackage{amssymb}
\usepackage{mathrsfs}
\usepackage{subfigure}
\usepackage{verbatim}
\usepackage{dsfont}
\usepackage{float}
\usepackage{tikz}
\usepackage{multirow}
\usepackage{etex}
\usepackage{tabularx,multirow,array,diagbox}
\usepackage[etex=true,export]{adjustbox}

\newcommand{\beq}{\begin{equation}}
\newcommand{\eeq}{\end{equation}}
\newcommand{\beql}{\begin{equation*}}
\newcommand{\eeql}{\end{equation*}}
\newcommand{\beqn}{\begin{eqnarray}}
\newcommand{\eeqn}{\end{eqnarray}}

\begin{document}
\title{Higher-order topological phases emerging from the Su-Schrieffer-Heeger stacking}
\author{Xun-Jiang Luo}
\author{Xiao-Hong Pan}
\affiliation{School of Physics and Institute for Quantum Science and Engineering, Huazhong University of Science and Technology, Wuhan, Hubei 430074, China}
\affiliation{Wuhan National High Magnetic Field Center and Hubei Key Laboratory of Gravitation and Quantum Physics, Wuhan, Hubei 430074, China}
\author{Chao-Xing Liu}
\affiliation{Department of Physics, the Pennsylvania State University, University Park, PA, 16802, US}
\author{Xin Liu}
\email{phyliuxin@hust.edu.cn}
\affiliation{School of Physics and Institute for Quantum Science and Engineering, Huazhong University of Science and Technology, Wuhan, Hubei 430074, China}
\affiliation{Wuhan National High Magnetic Field Center and Hubei Key Laboratory of Gravitation and Quantum Physics, Wuhan, Hubei 430074, China}
\begin{abstract}
In this work, we develop a systematical approach of constructing and classifying the model Hamiltonians for two-dimensional (2D) higher-order topological phase with corner zero energy states (CZESs). Our approach is based on the direct construction of analytical solution of the CZESs in a series of 2D systems that stack the 1D extended Su-Schrieffer-Heeger (SSH) model, two copies of the original SSH model,  along two orthogonal directions. Fascinatingly, our approach not only gives the celebrated Benalcazar-Bernevig-Hughes and 2D SSH models but also reveals a novel model and we refer it to crossed 2D SSH model.  Although these three models exhibit completely different bulk topology, we find that the CZESs can be universally characterized by edge winding number for 1D edge states, attributing to their unified Hamiltonian construction form and edge topology.  Remarkably, our principle of obtaining CZESs can be readily generalized to arbitrary dimension and superconducting systems. Thus, our work sheds new light on the theoretical understanding of the higher-order topological phase and paves the way to looking for higher-order topological insulators and superconductors. 
\end{abstract}
\maketitle

\textit{Introduction -}
Over the past few years, the concept of topological phases has been generalized to higher-order~\cite{Benalcazar2017a,Benalcazar2017,Song2017,Langbehn2017}, which has been extensively studied in electronic~\cite{Schindler2018,Geier2018,Schindler2018a,Khalaf2018}, bosonic~\cite{Xie2018,Serra-Garcia2018,Xie2019,Chen2019c,Ni2019,Fan2019,Xue2019}, Floquet~\cite{Rodriguez-Vega2019,Peng2019a,
Peng2020,Hu2020,Huang2020},  non-Hermitian~\cite{Liu2019,Zhang2019d,Luo2019,Edvardsson2019,Kawabata2020} and quasicrystal systems~\cite{Varjas2019,Chen2020a,Hua2020,Spurrier2020,Lv2021}. Especially in condensed matter system, the higher-order topological insulators ~\cite{Wang2020a,Li2020a,Liu2021,Li2021,Zhao2021} and superconductors~\cite{Hsu2020,Kheirkhah2020,Zhang2021,Luo2021a,Ghosh2021}, featuring  corner or hinge states, have
been attracting increasing attentions. Generally speaking, the
corner states of higher-order topological phase with additional chiral or particle-hole symmetry, will appear at
the center of their energy spectrum, namely zero energy.
Particularly in the superconducting system with intrinsic particle-hole symmetry, the exact zero energy corner
states, dubbed as Majorana corner states, follow non-Abelian braiding statistics
 and allow the implementation of topological quantum computation~\cite{Nayak2008}. Recently, it has been shown that the corner zero energy states (CZESs) in electronic system  also present nontrivial braiding properties \cite{Wu2020b}.

The CZEs have been studied in various higher-order topological systems~\cite{Yan2018,Hsu2018,Wang2018,Sheng2019,Pan2019,Zhu2019,Volpez2019,
Zhang2019,Ren2020,Wu2020a,Wu2020,Chen2020,Chen2021a}. However, the established topological invariants characterization of CZESs is usually case by case. For example,  the well-known Benalcazar-Bernevig-Hughes (BBH)~\cite{Benalcazar2017a} and 2D SSH models~\cite{Liu2017} are two paradigms featuring the CZEs, which are characterized by the quantized quadruple moment~\cite{Kang2019, Wheeler2019} and bulk polarization~\cite{Liu2017}, respectively. Although exhibiting different topological characterization, the BBH and 2D SSH models are both constructed from stacking the extended SSH models, two copies of the original SSH model~\cite{Su1979}. This motivates us to systematically investigate the 2D systems of stacking the 1D extended SSH models along different directions. For these systems, two open questions are urgent to be answered. Firstly, do these systems always support the CZESs? Secondly, are there general topological characterizations for the obtained CZESs?

\begin{figure}
\centering
\includegraphics[width=1\columnwidth]{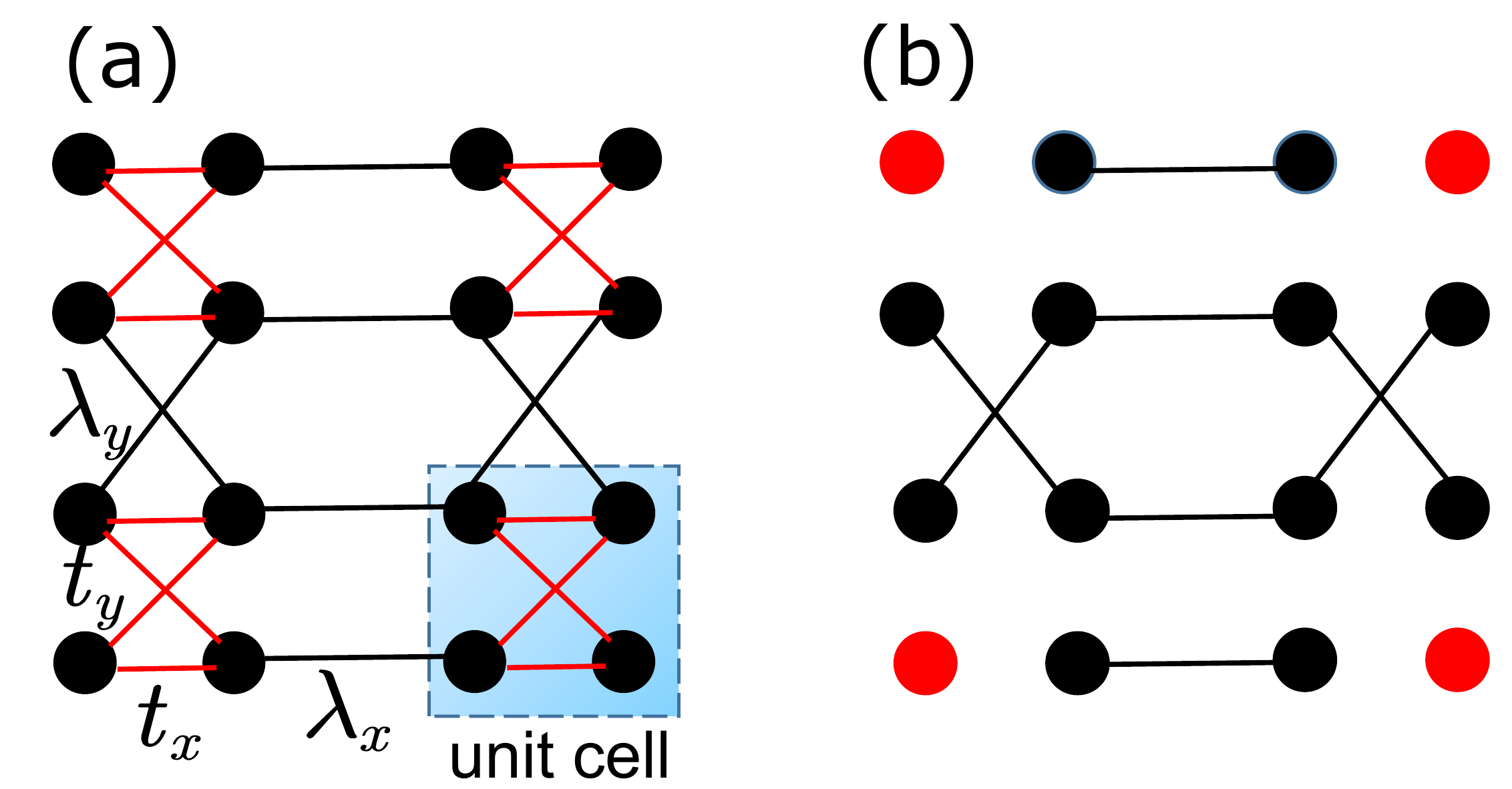}
\caption{  Schematic diagram for the lattice hoppings of crossed 2D SSH model. (a) Red and black bonds represent the intracellular and intercellular hoppings, respectively. (b)  The limit case $t_{x,y}=0$ in (a).}
\label{CSSH}
\end{figure}

In this work, we establish a general analytical theory to provide the CZESs existing condition in the systems, stacking the 1D extended SSH model along two orthogonal directions. This condition not only naturally presents the BBH and 2D SSH models, but also leads to a novel model displaying second-order topology. We can visually distinguish this new model from the BBH and 2D SSH models by its crossed hoppings along y-direction (Fig.\ref{CSSH}(a)). We thus call it crossed 2D SSH model. Although exhibiting completely different bulk topology, we find that the CZESs in the BBH, 2D SSH and 2D crossed SSH models can be uniformly characterized by the edge winding number for 1D edge states,  which elucidates the unified edge-corner correspondence~\cite{Ezawa2020,Trifunovic2020,Hu2021}. Moreover, Our theory can be easily generalized to 3D systems and predicts the mass of 3D higher-order topological phases.

\textit{Constructing corner zero energy states-}
Generally, we consider the 2D model Hamiltonian written as
\beqn
&H(\bm k)=\sum_{s=x,y}h_s(k_s),\nonumber\\
&h_{s}(k_s)=M_{s}(k_{s})\Gamma_{s}^{a}+\lambda_{s}\sin k_s\Gamma_{s}^{b},
\label{BH}
\eeqn
where $M_s(k_s)$ is defined as $t_{s}+\lambda_s\cos k_s$ and $\Gamma^{a(b)}_{s}$, belonging to 15 traceless $4\times 4$ Dirac matrices, satisfy the anti-commutation relation $\{\Gamma^a_s,\Gamma^b_s\} = 0$. Consequently, each 1D Hamiltonian $h_s$   respects chiral symmetry  $C_s = i\Gamma_s^a \Gamma_s^b$, with $C_s^2=1$. As each Dirac matrix has two-fold degenerate eigenvalue, the three Dirac matrices $\{\Gamma_s^{a}, \Gamma_s^{b}, C_s \}$ form the reducible representation of SU(2) Lie algebra: $h_{s}(k_s)$ can be considered as the direct sum of two copies of SSH model. Accordingly,  the topology of $h_s$ is determined by the winding number $\nu_s$ of the vector $(M_s,\lambda_s\sin k_s)$ around origin point~\cite{supp}. Then the topologically nontrivial phase is constrained in the region $|t_{s}|<|\lambda_s|$, corresponding to $\nu_s=1$. Taking $s=x$ for example, each end exists two end zero states in topologically nontrivial region and their wave functions can be obtained by solving the equation~\cite{supp}
\beqn\label{SSH-x}
h_x(x) |\Psi(x)\rangle=0,
\eeqn
with $h_x(x)$ the real space Hamiltonian. We find that the end zero states are the eigenstates of $C_x$ with eigenvalue $z_x$,  and the end states labelled by  $z_x=-1$ and $z_x=1$ are  localized at left and right ends, respectively.
 Consequently, the 1D end zero states wave function can be generically written as
\beqn\label{SL-x}
|\Psi_{z_x}(x)\rangle=f_{z_x}(x)|\psi_{z_x}\rangle.
\label{es}
\eeqn
where  $f_{-(+)}(x)$ is the spatial wave function localized at left (right) end and the spinor $|\psi_{z_x}\rangle$ satisfies $C_x|\psi_{z_x}\rangle=z_x|\psi_{z_x}\rangle$.  


\begin{figure}
\centering
\includegraphics[width=1\columnwidth]{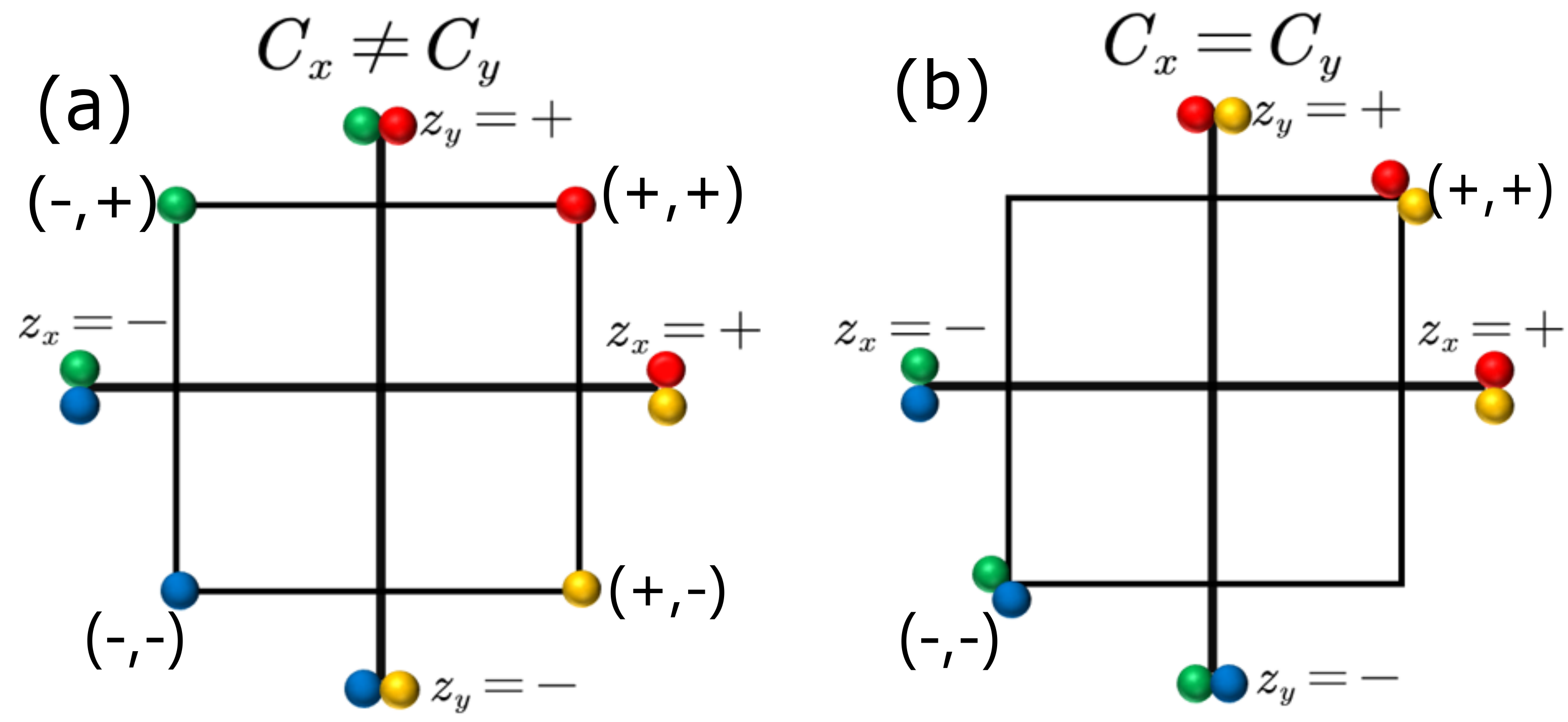}
\caption{  (a)(b)Schematic diagram of the end zero states and CZESs spatial  distribution under different situations. The black square frame denotes the boundary of the 2D system described by Hamiltonian $H(\bm k)$. The horizontal and perpendicular lines correspond to the 1D systems described by Hamiltonians $h_x$ and $h_y$, respectively. The four-color balls are the four common eigenstates of operators $C_x$ and $C_y$, denoting the end zero states or CZESs, labelled by eigenvalues ($z_x,z_y$). The relation between the end zero states and CZESs can be guided by the color of the balls. }
\label{cs}
\end{figure}

Similarly, for the 1D Hamiltonian $h_y$, we have
\beqn
&&h_y(y) |\Psi_{z_y}(y)\rangle = 0, \nonumber \\
&&|\Psi_{z_y}(y)\rangle  = g_{z_y}(y)|\psi_{z_y}(y)\rangle ,
\label{yes}
\eeqn
To understand above solution
visually, the end zero states of $h_x$ and $h_y$ are schematically denoted by the color balls in Figs.~\ref{cs}(a) and (b). However, we note that only the nontrivial topology of both $h_x$ and $h_y$ can not guarantee the existence of CZESs for 2D Hamiltonian $H$. This can be best exemplified by the well-known Bernevig-Hughes-Zhang model~\cite{Bernevig2006,BHZ}, in which the nontrivial topology of $h_x$ and $h_y$ give gapless edge states but without corner states. Remarkably, we find that  the CZEs can be obtained  when additional general condition,  namely $[C_x,C_y]_{-}=0$ is satisfied. Under this condition,  operators $C_x$ and $ C_y$ have four common eigenstates $|\psi_{(z_x,z_y)}\rangle$, labelled by their eigenvalues $(z_x,z_y)$, with $(z_x,z_y)\in\{(+,+),(+,-),(-,+),(-,-)\}$. Then we can construct the 2D wave function 
\beqn
\label{cczs}
|\Psi_{(z_x,z_y)}(\bm r)\rangle=f_{z_x}(x)g_{z_y}(y)|\psi_{(z_x,z_y)}\rangle.
\eeqn
It is easy to see that
\beqn
h_x(x) |\Psi_{(z_x,z_y)}(\bm r)\rangle=0,h_y(y) |\Psi_{(z_x,z_y)}(\bm r)\rangle=0,
\eeqn
resulting in $H(\bm r) |\Psi_{(z_x,z_y)}(\bm r)\rangle=0$. Obviously, the state $|\Psi_{(z_x,z_y)}(\bm r)\rangle$ exponentially decays along both $x,y$ directions, which indicates that it is a CZES for the 2D system. Thus, we can conclude that $H(\bm r)$ hosts four CZESs when $\nu_{x,y}=1$ and  $[C_x,C_y]_{-}=0$. 

{\it Classification-}
Explicitly, the condition $[C_x,C_y]_{-}=0$ can be divided into two situations, namely (a): $C_x \neq C_y$ and (b): $C_x = C_y$. Here, we do not distinguish the equivalent  cases $C_x=C_y$ and $C_x=-C_y$.  For situation (a), the four common eigenstates are labeled by $(z_x,z_y)=\{(+,+),(+,-),(-,+),(-,-)\}$. Thus, the corresponding four CZESs, schematically distinguished  by the red, yellow, green and blue balls in Fig.~\ref{cs}(a),  are localized at each corner according to Eq.~\ref{cczs}. For situation (b), because of $z_x=z_y$, the four common eigenstates of $C_x$ and $C_y$ are labeled by $(z_x,z_y)=\{(+,+),(+,+),(-,-),(-,-)\}$. As a result, the four corresponding CZESs are localized at the diagonal corners,  shown in Fig.~\ref{cs}(b). Notably,  up to now our analysis is general and the  specific form of $H(\bm k)$ has not been given. However, the topological property of $H(\bm k)$  deeply depend on the given form.  In the following, we perform classification of $H(\bm k)$ under the condition $[C_x,C_y]=0$.

As any two Dirac matrices either commute or anti-commute to each other, the second-order topological phase described by $H(k)$ can be further classified by the commutation relations between $\Gamma_x^{a,b}$ and $\Gamma_y^{a,b}$ under the condition $[C_x,C_y]=0$, or equivalently $[i\Gamma_x^a\Gamma_x^b,i\Gamma^a_y\Gamma_y^b] = 0$. 
 It is straightforwardly to show that there exist four  inequivalent cases with the commutation relations 
\beqn
\label{cases}
&&(\text{i}): \{\Gamma_x^a,\Gamma_y^{a,b}\}=0, \{\Gamma_x^b,\Gamma_{y}^{a,b}\}=0;\nonumber\\
&&(\text{ii}): [\Gamma_x^a,\Gamma_y^{a,b}]=0, [\Gamma_x^b,\Gamma_y^{a,b}]=0;\nonumber\\
&&(\text{iii}): [\Gamma_x^a,\Gamma_y^{a,b}]=0, \{\Gamma_x^b,\Gamma_y^{a,b}\}=0;\nonumber\\
&&(\text{iv}): [\Gamma_x^a,\Gamma_y^{a}]=0,  \{\Gamma_x^b,\Gamma_y^{b}\}=0\nonumber\\
&&\quad\quad \{\Gamma_x^a,\Gamma_y^{b}\}=0, [\Gamma_x^b,\Gamma_y^{b}]=0
\label{cl}
\eeqn
Considering concrete representation of the Dirac matrices, we find that  situations   $C_x\neq C_y$ and $C_x = C_y$ correspond to the cases (i-iv) and (iv), respectively~\cite{supp}. On the other hand, it can be readily verified that $H(\bm k)$ has bulk chiral symmetry $\mathcal{C}$ with $\{\mathcal{C},H(\bm k)\}=0$ for all the cases. Concretely,  $\mathcal{C}=C_xC_y$ and $\mathcal{C}=C_x$ for cases (i-ii) and cases (iii-iv), respectively. Since the CZESs are labelled by eigenvalues ($z_x,z_y$), the CZEs are  the eigenstate of $\mathcal{C}$, with eigenvalue $z=z_xz_y$ or $z=z_x$. With this property, the CZESs labelled by the same eigenvalue of operator $\mathcal{C}$ can not be coupled by the perturbations preserving the bulk chiral symmetry~\cite{per}, which allows a Z topological classification for the CZESs of second-order topological insulator phase.

In case (i), matrices $\{\Gamma_{x}^{a,b},\Gamma_{y}^{a,b}\}$ anti-commute with each other, corresponding to the BBH model. 
In case (ii), $h_x$ and $h_y$ commute with each other, corresponding to the 2D SSH model. Remarkably, the commutation relations in the (iii) and (iv) predict two unprecedented models. Case (iii) corresponds to the crossed 2D SSH model (Fig.~\ref{CSSH}). In case (iv),  we find that the CZESs always coexist with the edge flat band~\cite{supp}, which brings the difficulty to identify and characterize the CZESs. In the following, we focus on the crossed 2D SSH model.

\begin{figure}
\centering
\includegraphics[width=1\columnwidth]{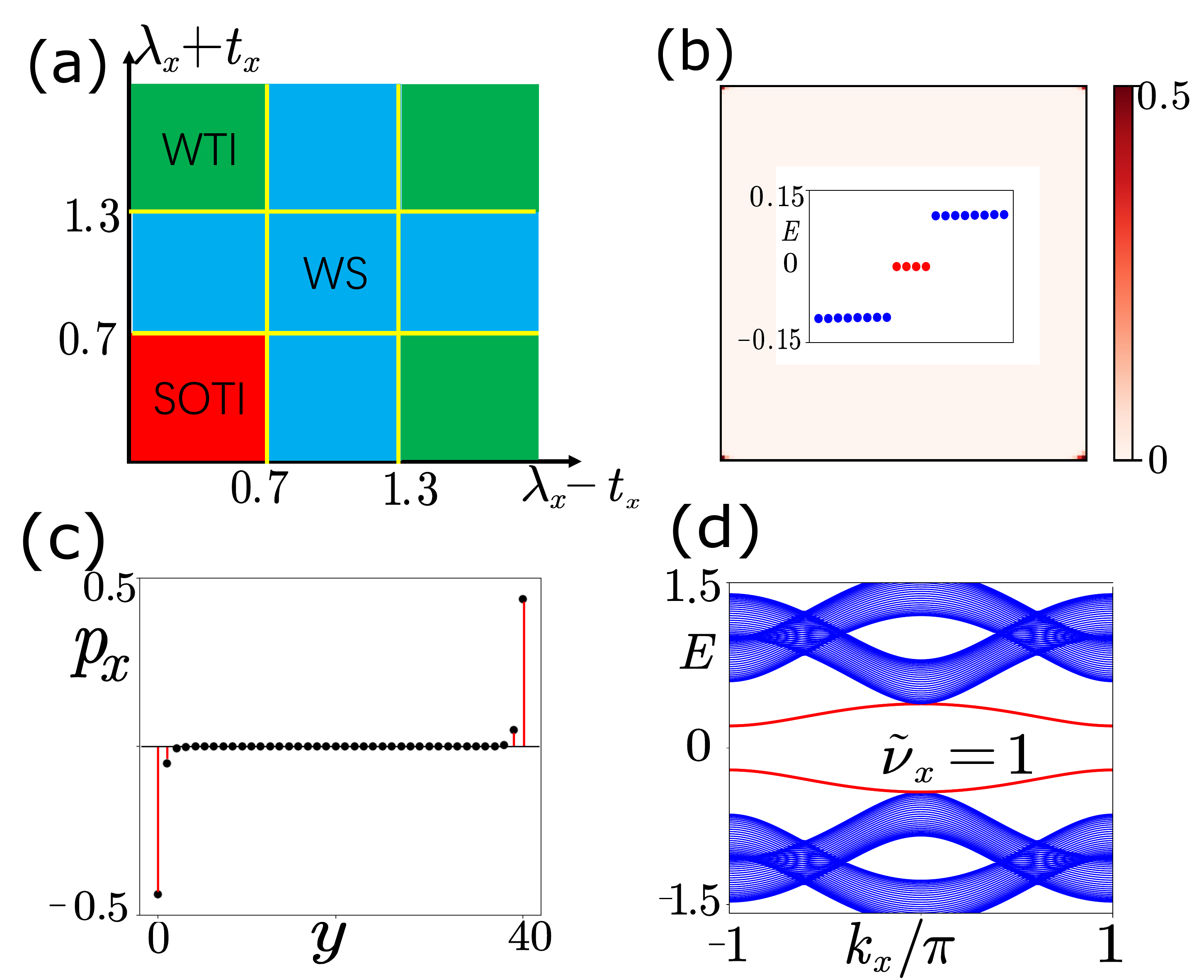}
\caption{(a) The phase digram of the second-order topological phase in 2D crossed model. The bulk phase transitions, represented by the yellow line, divide the bulk states into weak topological insulator (WTI), Weyl semimetal (WS), or second-order topological insulator (SOTI) phases.   (b) Spatial distribution of the CZESs in the 2D crossed SSH model, the inset plots the eigenenergies close to zero.  (c) Numerical calculation of edge polarization $p_x$. (d) Energy dispersion of nanoribbon structure along $y$ direction.  The red bands denote the edge states.  Common parameters in (a)(c)(d) are taken with $t_x=0.1,\lambda_x=0.2,t_y=0.3,\lambda_y=1$.}
\label{hop}
\end{figure}

{\it Crossed 2D SSH model-}
Considering the concrete representation of the Dirac matrices, the Hamiltonian for case (iii)  can be written as
\beqn
&&\mathcal{H}(\bm k)=h_x(k_x)+h_y(k_y),\nonumber\\
&&h_x(k_x)=M_x(k_x)\tau_x\sigma_0+\lambda_x\sin k_x\tau_y\sigma_0,\nonumber\\
&&h_y(k_y)=M_y(k_y)\tau_x\sigma_x+\lambda_y\sin k_y\tau_x\sigma_y,
\label{CRSSH}
\eeqn
with $\tau,\sigma$ two sets of Pauli matrices. The corresponding lattice hopping of $\mathcal{H}(\bm k)$ is  schematically shown in Fig.~\ref{CSSH}(a), which has dimerized hopping in $x$-direction as the 1D SSH model and crossed hopping in $y$-direction. Therefore, we refer to this model as the crossed 2D SSH model.  In Fig.~\ref{CSSH}(b), the isloated atoms at the corner in the limit case $t_{x,y}=0$ correspond to the CZESs.

To study the bulk phase and band structures of $\mathcal{H}(\bm k)$, 
we simplify  $\mathcal{H}(\bm k)$ as
\beqn\label{CSSH-2}
\mathcal{H}(\bm k) =M_x(k_x)\tau_x\sigma_0+\lambda_x\sin k_x\tau_y\sigma_0+E_y\tau_x \sigma_\varphi,
\eeqn
with $\sigma_\varphi=\cos\varphi\sigma_x+\sin\varphi\sigma_y$, $\cos\varphi=M_y/E_y$, $E_y=\sqrt{M_y^2+(\lambda_y\sin k_y)^2}$. In the eigenbasis of  $\sigma_{\varphi}$ ($\sigma_{\varphi}= \pm 1$), $\mathcal{H}(\bm k) $ is block-diagonal and the two blocks Hamiltonian can be written as
\beqn\label{SSH-2}
h_{\pm}(\bm k)=(M_x\pm E_y)\tau_x+\lambda_x\sin k_x\tau_y,
\eeqn
with $\pm$ corresponding to the eigenvalues of $\sigma_{\varphi}$.
 As a result, we can reveal the 2D bulk spectrum of $\mathcal{H}(\bm k)$ in Eq.~\eqref{CSSH-2} through the spectrum of $h_{\pm}(\bm k)$, which can be considered as the 1D SSH model along $k_x$-direction with $k_y$-dependent hopping amplitude $t_x\pm E_y$. Given $k_y$, the topology of 1D SSH Hamiltonians $h_{\pm}(\bm k)$ is characterized by the quantized Berry phase $\alpha_{\pm} (k_y)$ of occupied states. We classify the bulk states of Eq.~\eqref{CSSH-2} into three phases based on the first-order topological band theory: when $\alpha_{+}(k_y)$ and $\alpha_{-}(k_y)$ are both quantized to $\pi$ over all the range of $k_y$, the SSH models in Eq.~\eqref{SSH-2} are fully gapped so that the bulk energy spectrum of $\mathcal{H}(\bm k)$ in Eq.~\eqref{CSSH-2} is also fully gapped. The system can be viewed as the pile-up of 1D topologically non-trivial SSH model,  which is weak topological insulator with flat edge band~\cite{supp}; when $\alpha_{+}(k_y)$ and $\alpha_{-}(k_y)$ are both quantized to 0 over all the range of $k_y$, the SSH models in Eq.~\eqref{SSH-2} and the bulk energy spectrum of $\mathcal{H}(\bm k)$ in Eq.~\eqref{CSSH-2} are also fully gapped. The system is a trivial insulator; when $\alpha_{+}(k_y)$ or $\alpha_{-}(k_y)$ has a transition with varying $k_y$, the SSH models in Eq.~\eqref{SSH-2} and the bulk energy spectrum of $\mathcal{H}(\bm k)$ in Eq.~\eqref{CSSH-2} close their gaps at certain $k_y$ along high symmetry line $k_x=0$ or $\pi$. 
\begin{table}[htb]
\centering
\setlength\tabcolsep{2.5pt}
\renewcommand{\arraystretch}{1.1}
\caption{ Topological characterization of the CZESs by various topological invariants. Here, $\tilde{\nu}$ denotes the edge winding number of 1D edge states along $x$-direction.}
\begin{tabular}{|c|c|c|c|c|}
\hline
   \diagbox{Case}{invariant} & $(p_x^{\nu_y}, p_y^{\nu_x})$ & $(P_x,P_y)$ & $\quad Q_{xy}\quad$&$\quad\tilde{\nu_x}\quad$ \\
  \hline
  (i)(BBH) & \checkmark & $\bm{\times}$ & \checkmark &\checkmark \\
  \hline
   (ii)(2D SSH) & $\bm{\times}$ &  \checkmark & $\bm{\times}$& \checkmark \\
  \hline
  (iii) &  $\bm{\times}$ &  $\bm{\times}$ & \checkmark&\checkmark\\
  \hline
\end{tabular}
\label{tab}
\end{table}
The system becomes mirror symmetry ($\mathcal{M}_x=\tau_x\sigma_0$) protected Weyl semimetal~\cite{MS,supp}. Having clarified the bulk phase, we plot the bulk phase diagram in Fig.~\ref{hop}(a) under the parameters $\lambda_y=1,t_y=0.3$ and $0<t_x<\lambda_x$. In this CZESs existing parameters region,  we find that the bulk states can be  divided into second-order topological insulator (the CZESs shown in Fig.~\ref{hop}(b)), weak topological insulator and Wely semimetal through the creation or annihilation of the Wely points, represented by the yellow line in Fig.~\ref{hop}(a). The absence of bulk-corner correspondence  implies that the CZESs may not origin from the bulk topology.

 On the other hand, for characterizing the CZESs, several bulk topological invariants have been established, including nested Wilson loop ($p_x^{\nu_y}, p_y^{\nu_x}$)~\cite{Benalcazar2017a,Benalcazar2017}, bulk polarization ($P_x,P_y$)~\cite{Liu2017,Benalcazar2019}, quadrupole moment $Q_{xy}$~\cite{Kang2019,Wheeler2019}. Here, we test the applicability of characterizing the  2D crossed SSH model  by these topological invariants and take the BBH and 2D SSH models as comparison. We find that the nested Wilson loop and polarization topological characterizations only apply to BBH and 2D SSH model, respectively~\cite{supp}. Moreover, the quadrupole moment topological characterization apply to both BBH and crossed 2D SSH models~\cite{supp}. However distinct from the BBH model with edge polarization~\cite{Benalcazar2017a,Benalcazar2017} $p_x^{\text{edge}}=p_y^{\text{edge}}=0.5$, the crossed 2D SSH model exhibits nontrivial edge polarization only along $x$-direction~\cite{supp} and $p_x^{\text{edge}}=0.5$ is shown in Fig.~\ref{hop}(c).  Thus, the crossed 2D SSH model provides a paradigm of type-II quadrupole topological insulator~\cite{Yang2020a}, featuring nonzero quadrupole moment and one direction edge polarization. Although absence of  unified characterization by these bulk topological invariants, we find that the CZESs in BBH, 2D SSH and 2D crossed SSH models can be uniformly characterized by nonzero winding number for 1D edge states \cite{supp}.  We  take the crossed 2D SSH model for example following. 

Note that the in gap edge state (red curves in Fig.~\ref{hop}(d)), corresponding to the edge-localized states, extend over the whole 1D Brillouin zone. Thus, these edge states can be described by truly 1D lattice Hamiltonian, which is essential to define edge winding number for 1D edge states unambiguously. Directly, the wave function of the edge states can be obtained by solving the equation $\mathcal{H}(k_x,y)|\Psi(k_x,y)\rangle=E(k_x)|\Psi(k_x,y)\rangle$. Because of $[h_x,C_y]_{-}=0$,  the edge state $\Psi(k_x,y)\rangle$ are the common eigenstate of $C_y$ and $h_x$, showing as
\beqn
&|\Psi(k_x,y)\rangle=g_{+}(y)P|\Phi(k_x)\rangle,\nonumber\\
&h_x(k_x)|\Phi(k_x)\rangle=E_x(k_x)|\Phi(k_x)\rangle,
\label{ESC}
\eeqn
with $E_x(k_x)=\sqrt{M_x^2+(\lambda_x\sin k_x)^2}$ and the edge projection operator $P=(1+ C_y)/2$~\cite{Khalaf2018a,Roberts2020}.  The edge Hamiltonian can be obtained by projecting $h_x$ to the subspace defined by $P$, leading to the edge Hamiltonian
\beqn
\tilde{h}_x(k_x)=M_x(k_x)\tilde{\tau}_x+\lambda_x\sin k_x\tilde{\tau}_y.
\label{EH}
\eeqn
with $\tilde{\tau}$ acting the subspace where $C_y=1$. Obviously, $\tilde{h}_x(k_x)$ behaves as the SSH model and is topologically nontrivial when $\nu_x=1$. On the other hand, the existence of edge states  depend on the condition  that $h_y$ is topologically nontrivial, namely $\nu_y=1$. Thus, the edge winding number $\tilde{\nu}_x=1$ defined by the occupied states of $\tilde{h}_x$ can precisely characterize the CZESs existing condition $\nu_{x,y}=1$. Remarkably, the commutation relation $[h_x,C_y]=0$ is the main reason for the existence of well defined winding number of 1D edge states. It can be readily that $[h_x,C_y]=0$ also holds in the BBH and 2D SSH model according to Eq.~\eqref{cases}. Thus, the edge winding number characterization also applies to  the BBH and 2D SSH models owing to the unified commutation relation $[h_x,C_y]=0$, which also reflects the CZESs existing condition $[C_x,C_y]=0$.

{\it Higher dimensional generalizations-}
 Our principle of constructing the CZESs can be easily generalized to arbitrary dimension~\cite{supp}. Here,
we consider 3D eight bands Hamiltonian
\beqn
&&H(\bm k)=\sum_{s=x,y,z}h_s(k_s),\nonumber\\
&&h_{s}(k)=(t_s+\lambda_s\cos k_s)\Gamma_{as}^{8}+\lambda_s\sin k_s\Gamma_{bs}^{8},
\eeqn
where $8 \times 8$ Gamma matrices $\Gamma_{as,bs}^{8}$~\cite{supp} anti-commute with each other and the chiral symmetry  of $h_s$ is given by $C_s=i\Gamma_{as}^{8}\Gamma_{bs}^{8}$. Similarly, $h_s$ can be deemed as the direct sum of four copies of SSH model and then $h_s$ exists four end zero states at each end. Remarkably, when $8\times 8$ matrices $\{C_x,C_y,C_z\}$ commute with each other, they have eight common eigenstates. Correspondingly, eight CZESs of 3D Hamiltonian $H$ can be constructed according our general principle. However, to determine the topological property of $H(\bm k)$, we need to specify the commutation relations between all the Gamma matrices.  On the other hand, our 2D classification indicates that there are four types commutation relations between matrices $\{\Gamma_{as}^8,\Gamma_{bs}^8,\Gamma_{as^{'}}^8,\Gamma_{bs^{'}}^8\}$ under the condition $[C_{s},C_{s^{'}}]_{-}=0$, where  $\{s,s^{'}\}\in\{x,y\},\{x,z\},\{y,z\}$. Thus, classifying $H(\bm k)$ can  predict 64 models featuring CZESs when do not distinguish the equivalent status between different directions.  A typical example is the topological octupole insulator model~\cite{Benalcazar2017a,Benalcazar2017,Bao2019}, in which all the Gamma matrices anti-commute with each other. We study other predicted models featuring the CZESs in our future work.

{\it Discussion and Conclusion-}
It is noted that our theory of obtaining the CZESs can readily be generalized to the superconducting system by requiring an additional particle-hole symmetry. Some higher-order topological superconductors can be predicted~\cite{supp}. In view of the experiment realization of the BBH and 2D SSH models~\cite{Serra-Garcia2018,Serra-Garcia2019,Imhof2018,Mittal2019,Xie2019,Chen2019c,Zheng2019,Qi2020}, we believe that the crossed 2D SSH model can also be realized in various artificial lattice systems. It is also worth emphasizing that our theory of obtaining the CZESs can be easily generalized to obtain the hinge states with analytical solutions. Thus in our theoretical framework, we can systematically  construct arbitrary order topological insulators and superconductors in arbitrary dimension by directly constructing the analytical solution of boundary states, which is left as an independent work.

In summary, we provide a general analytical theory to study the higher-order topological phase emerging from SSH stacking. Our theory not only gives the well-known BBH and 2D SSH models, but also predicts the crossed 2D SSH model. We establish the unified topological characterizing of these three models. Our work provides a broad venue to looking for higher-order topological phases in arbitrary dimension.

%


\clearpage
\begin{appendix}
\begin{widetext}
\begin{center}
\begin{Large}
\textbf{Supplemental Materials}
\end{Large}
\end{center}

\section{Dirac matrices and their generalization}
Starting from three anti-commuting Pauli matrices $\sigma_{x,y,z}$ and $ 2\times 2$ identity matrix $\sigma_0$, the 16 Dirac matrices $\sigma_{i}\otimes \sigma_j(\sigma_i\sigma_j)$ can be obtained through their direct product, with $i=j=x,y,z,0$. Besides $4\times 4$ identify matrix, the other 15 Dirac matrices are traceless and they square to identify. For the 15 traceless Dirac matrices, the five of them anti-commuting with each other. Without loss of generality, we can choose the five anti-commuting matrices as
\beqn
\Gamma_1^{4}=\sigma_z\sigma_x,\Gamma_2^{4}=\sigma_z\sigma_y,\Gamma_3^{4}=\sigma_z\sigma_z,\Gamma_4^{4}=\sigma_x\sigma_0,\Gamma_5^{4}=\sigma_y\sigma_0.
\eeqn
Other 10 traceless Dirac matrices can be generated by $\Gamma_{mn}^{4}=\frac{1}{2i}[\Gamma_m^{4},\Gamma_n^{4}]$, with $m=n=1,2,3,4,5$. Generalizing to higher dimension, the direct product of arbitrary $d$ sets Pauli matrices can generate $4^{d}$ Gamma matrices $\sigma_i\cdots\sigma_j\cdots\sigma_k$ with dimension $2^{d}$ and they square to identify. In these $4^d$ Gamma matrices,  $2d+1$ matrices anti-commuting with each other, forming complex Clifford algebra. Generally, the $2d+1$ anti-commuting matrices can be obtained through the iteration from $2d-1$ anti-commuting Gamma matrices with dimension $2^{d-1}\times 2^{d-1}$
\beqn
\Gamma_{1,2,\cdots,2d-1}^{2^d}=\sigma_z\otimes \Gamma_{1,2,\cdots,2d-1}^{2^{d-1}}, \Gamma_{2d}^{2^d}=\sigma_x\otimes I^{2^{d-1}}, \Gamma_{2d+1}^{2^d}=\sigma_y\otimes I^{2^{d-1}},
\eeqn
where $ I^{2^{d-1}}$ denotes the $2^{d-1}\times 2^{d-1}$ identify matrix, $\Gamma_{1,2,\cdots,2d-1}^{2^{d-1}}$ represents $2d-1$ anti-commuting Gamma matrices with dimension $2^{d-1}\times 2^{d-1}$.

\section{1D extended SSH model}
\label{wdwf}
In the momentum space, we consider the general model Hamiltonian in AIII symmetry class\cite{Schnyder2008,Ryu2010,Chiu2016}.
\beqn
h(k)=M(k)\Gamma_{a}^{2^d}+\lambda\sin k\Gamma_{b}^{2^d},
\label{sh}
\eeqn
where $M(k)=(t+\lambda\cos k)$, $\Gamma_{a,b}^{2^d}$ are $2^d\times 2^d$ Gamma matrices and satisfy $\{\Gamma_{a}^{2^d},\Gamma_{b}^{2^d}\}=0$. The chiral symmetry of $h$ can be written as $C=i\Gamma_{a}^{2^d}\Gamma_{b}^{2^d}$. It is noted that $h(k)$ is block-diagonal in certain basis and each $2\times 2$ block Hamiltonian behave as the SSH model. Thus, $h(k)$ can be generically deemed as the direct sum of $2^{d-1}$ copies of SSH model. In the following, we characterize the topology of $h(k)$ by topological invariant winding number.

The energy spectrum of $h$ is $E=\sqrt{(t+\lambda\cos k)^2+(\lambda\sin k)^2}$. For simplicity, $h$ can be normalized as
\beqn
\bar{h}=\cos\varphi\Gamma_{a}^{2^d}+\sin\varphi\Gamma_{b}^{2^d},
\eeqn
with $\cos\varphi=(t+\lambda\cos k)/E$.
With the dimension and symmetry class given, the topology of $\bar{h}$ is determined by the winding number
\beqn
\bar{\nu}&=&-\frac{1}{4i\pi }\int_{-\pi}^{\pi}\text{Tr}[C\bar{h}d\bar{h}]\\
&=&-\frac{1}{4i\pi }\int_{-\pi}^{\pi}\text{Tr}[(\cos\varphi\partial_{k}\cos\varphi+\sin\varphi\partial_{k}\sin\varphi)C+(\cos\varphi\partial_{k}\sin\varphi-\sin\varphi\partial_{k}\cos\varphi)C\Gamma_{a}^{2^d}\Gamma_{b}^{2^d}]\\
&=&\frac{2^{d}}{4\pi }\int_{-\pi}^{\pi}(\cos\varphi\partial_{k}\sin\varphi-\sin\varphi\partial_{k}\cos\varphi)\\
&=&\frac{2^{d}}{4\pi }\int_{-\pi}^{\pi}\partial_{k}\varphi.
\eeqn
In the parameter region $|t|<|\lambda|$, above integration yields topological invariant $\bar{\nu}=2^{d-1}$. Otherwise, $\bar{\nu}=0$. Owing to the bulk-boundary correspondence, the winding number $\bar{\nu}$ is associated with $2^{d-1}$ end zero states localized at each end under the open boundary condition. In the following, we solve the analytical wave function of these end zero states.

Considering the semi-infinite system ($r>0$) described by $h$, we solve the end zero states localized close to the end $r=0$. Directly, we expand the Hamiltonian $h$ at $k=0$ to second order of $k$ and replace $k\rightarrow -i\partial_{r}$. Then we have
\beqn
h(-i\partial_{r})=(m+\lambda/2\partial_{r}^2)\Gamma_{a}^{2^d}-i\lambda\partial_{r}\Gamma_{b}^{2^d},
\eeqn
with $m=t+\lambda$. Solving the eigen equation $h(-i\partial_{r})|\Phi_{\alpha}(r)\rangle=0$ gives rise to
\beqn
(m+\lambda/2\partial_{r}^2)\Gamma_{a}^{2^d}|\Phi_{\alpha}(r)\rangle-i\lambda\partial_{r}\Gamma_{b}^{2^d}|\Phi_{\alpha}(r)\rangle=0.
\eeqn
Multiplying both sides by $\Gamma_{a}^{2^d}$ gives
\beqn
(m+\lambda/2\partial_{r}^2)|\Phi_{\alpha}(r)\rangle=\lambda\partial_{r}C|\Phi_{\alpha}(r)\rangle.
\label{eq}
\eeqn
Obviously,  state $|\Phi_{\alpha}(r)\rangle$ should be the eigenstate of chiral operator $C$, namely $C|\Phi_{z}(r)\rangle=z|\Phi_{z}(r)\rangle$ with $z=\pm 1$. We set the trial wave function $|\Phi_{z}(r)\rangle=e^{\xi_{z} r}|\psi_{z}\rangle$, with $C|\psi_{z}\rangle=z|\psi_{z}\rangle$ and $\xi_{z}$ is a complex number. By inserting this ansatz solution into Eq.~\eqref{eq}, we have
\beqn
\lambda_s/2\xi_{z_s}^2-z_s\lambda_s\xi_s+m=0.
\eeqn
The two roots are $ \xi_{z}^{1,2}=\frac{z\lambda \pm \sqrt{\lambda^2-2m\lambda}}{\lambda}$. In the region $|t|<|\lambda|$, the real part of $ \xi_{z}^{1,2}$ are negative and positive when $z=-1$ and $z=1$, respectively. Under the boundary condition $|\Phi_{z}(0)\rangle=|\Phi_{z}(\infty)\rangle=0$, we can know that the wave function of end states are $|\Phi_{-}(r)\rangle=\mathcal{N}(e^{\xi_{-}^{1}r}-e^{\xi_{-}^{2}r})|\psi_{-}\rangle$, with the normalization factor $\mathcal{N}$. On the contrast, if we consider the semi-system $r<0$, then we will find that the end zero states should be the eigenstate of chiral operator $C$ with eigenvalue $z=1$. As a result, for a finite system with length $L$, the end zero states localized close to the end $r=0$ and $r=L$ are the eigenstates of chiral operator $C$ with eigenvalue $z=-1$ and $z=1$, respectively. For $2^{d}\times 2^{d}$ matrix $C$, there are $2^{d-1}$ eigenstates with eigenvalue $z=1$ and $z=-1$, respectively. Thus, there are $2^{d-1}$ end zero states localized at each end for $h(r)$. In the main text, we take $d=2$ and $d=3$, then there are two and four end zero states localized at each end, respectively. The spatial parts of the wave function for these end zero states are
\beqn
&f_{-}^{s}(r_s)=\mathcal{N}_s^{-}(e^{\xi_{-}^1r_s}-e^{\xi_{-}^2r_s}), f_{+}^{s}(r_s)=\mathcal{N}_s^{+}(e^{\xi_{+}^1(r_s-L_s)}-e^{\xi_{+}^2(r_s-L_s)}),
\eeqn
where $\mathcal{N}_s^{-}, \mathcal{N}_s^{+}$ are  the normalization factors,  index $s$ denotes the different directions.


\section{General principle of obtaining the CZESs}
Considering arbitrary $d$D Bloch Hamiltonian
\beqn
H(\bm k)=\sum_{s=1}^{d}h_s(k_s), \{C_s, h_s(k_s)\}=0,
\label{dD}
\eeqn
where $\bm k=(k_1,\cdots, k_d)$, $h_s$ belongs to the AIII symmetry class and respects the chiral symmetry $C_s$. Thus, 1D Hamiltonian $h_s$ has a Z topological classification~\cite{Schnyder2008,Ryu2010,Chiu2016}. When $h_s$ is topologically nontrivial characterized by the nonzero winding number, there are end zero states for this 1D system and their wave functions can be generically written as
\beqn
|\Phi^s_{z_s}(r_s)\rangle=f_{z_s}^{s}(r_s)|\psi^s_{z_s}\rangle,C_s|\psi^s_{z_s}\rangle=z_s |\psi^s_{z_s}\rangle,
\label{ezs}
\eeqn
where spinor $|\psi^s_{z_s}\rangle$ is the eigenstate of $C_s$ with eigenvalue $z_s=\pm 1$, scalar function $f_{z_s}^{s}(r_s)$ exponentially decays along $r_s$. Here, we have used the fact the end zero states always can be labeled by the eigenvalue of chiral symmetry $C_s$. Remarkably, when $|\psi^{1}_{z_1}\rangle=\cdots=|\psi^{d}_{z_d}\rangle=|\psi_{(z_1,\cdots,z_d)}\rangle$, we can construct the $d$D wave function
\beqn
|\Psi_{z_1,\cdots,z_d}(\bm r)\rangle=\prod_{s=1}^df_{z_s}^s(r_s)|\psi_{(z_1,\cdots,z_d)}\rangle.
\eeqn
It is easy to see that
\beqn
h_{s=1,\cdots,d}|\Psi_{z_1,\cdots,z_d}(\bm r)\rangle=0, H(-i\partial_{\bm r})|\Psi_{z_1,\cdots,z_d}(\bm r)\rangle=0.
\eeqn
Thus, the state $|\Psi_{z_1,\cdots,z_d}(\bm r)\rangle$ is the zero energy state of Hamiltonian $H(-i\partial_{\bm r})$. Obviously, state $|\Psi_{z_1,\cdots,z_d}(\bm r)\rangle$ exponentially  decays along all  directions. Therefore, it is localized at the corner of a $d$D  system and we obtain a CZESs.

Without loss of generality, we exemplify 1D Hamiltonian $h_{s}(k_{s})$ with the form considered in Eq.~\eqref{sh}. Explicitly, the considered Hamiltonian can be written as
\beqn
H(\bm k)=\sum_{s=1}^dh_s(k_s), h_s(k_s)=M_s(k_s)\Gamma_{as}^{2^d}+\lambda_s\sin k_s\Gamma_{bs}^{2^d}.
\label{ad}
\eeqn
Under the condition $|t_s|<|\lambda_s|$, we have shown that $h_s$ is topologically nontrivial and the bulk topology is characterized by the winding numbers $\bar{\nu}_s=2^{d-1}$.
Taking the open boundary condition of $k_s$, $h_s$ hosts $2^{d-1}$ end zero states localized at the ends $r_s=0$ and $r_s=L_s$, respectively. The wave function of the end zero states can be written as $|\Phi_{z_s}^s(r_s)\rangle=f_{z_s}^s(r_s)|\psi_{z_s}^s\rangle$ with $C_s|\psi_{z_s}^s\rangle=z_s|\psi_{z_s}^s\rangle$.

When $2^d\times 2^d$ matrices $\{C_1,\cdots,C_m,\cdots, C_d\}$ commute with each other, they have $2^d$ common eigenstates labelled by their eigenvalue $(z_1,\cdots,z_m\cdots,z_d)$. Correspondingly, we obtain $2^d$ CZESs with the wave function
\beqn
|\Psi_{z_1,\cdots,z_d}(\bm r)\rangle=\prod_{s=1}^df_{z_s}^s(r_s)|\psi_{(z_1,\cdots,z_m\cdots,z_d)}\rangle,
\eeqn
with $C_m|\psi_{(z_1,\cdots,z_m\cdots,z_d)}\rangle=z_m|\psi_{(z_1,\cdots,z_m\cdots,z_d)}\rangle$.
It can be readily verified that $h_{s=1,2\cdots,d}|\Psi_{z_1,\cdots,z_d}(\bm r)\rangle=0$, giving rise to $H(-i\partial_{\bm r})|\Psi_{z_1,\cdots,z_d}(\bm r)\rangle=0$. It is noted that the commutation relations between $\Gamma_{js}^{2^{d}}$ and $\Gamma_{j^{'}s^{'}}^{2^{d}}$ have not been given still, with $s,s^{'}\in\{1,\cdots,d\}$, $j,j^{'}=a,b$. However, these commutation relations will determine the topology property of ${H}(\bm k)$ unambiguously. Under the condition $[C_s,C_{s^{'}}]=0$, 2D classification in the main text indicates there are four types commutation relations between $\Gamma_{js}^{2^{d}}$ and $\Gamma_{j^{'}s^{'}}^{2^{d}}$, namely
\beqn
\label{cases-2}
&&(\text{i}): \{\Gamma_{as}^{2^d},\Gamma_{as^{'},bs^{'}}^{2^d}\}=0, \{\Gamma_{bs}^{2^d},\Gamma_{as^{'},bs^{'}}^{2^d}\}=0,\nonumber\\
&&(\text{ii}): [\Gamma_{as}^{2^d},\Gamma_{as^{'},bs^{'}}^{2^d}]=0, [\Gamma_{bs}^{2^d},\Gamma_{as^{'},bs^{'}}^{b^d}]=0,\nonumber\\
&&(\text{iii}): [\Gamma_{as}^{2^d},\Gamma_{as^{'},bs^{'}}^{2^d}]=0, \{\Gamma_{bs}^{2^d},\Gamma_{as^{'},bs^{'}}^{b^d}\}=0,\nonumber\\
&&(\text{iv}):  [\Gamma_{as}^{2^d},\Gamma_{as^{'}}^{2^d}]=0,  \{\Gamma_{as^{'}}^{2^d},\Gamma_{bs^{'}}^{2^d}\}=0,\nonumber\nonumber\\
&&\quad\quad \{\Gamma_{as}^{2^d},\Gamma_{bs^{'}}^{2^d}\}=0, [\Gamma_{as^{'}}^{2^d},\Gamma_{bs^{'}}^{2^d}]=0.
\label{case}
\eeqn
Therefore, there are $4\times 4^{2}\times\cdots\times 4^{d-1}$ types commutation relations for ${H}(\bm k)$ when assign all the commutation relations between $\Gamma_{js}^{2^{d}}$ and $\Gamma_{j^{'}s^{'}}^{2^{d}}$, with $s,s^{'}\in\{1,\cdots,d\}$, $j,j^{'}=a,b$. Notably, here we do not distinguish the equivalent status between different directions.
Once the commutation relations between all these Gamma matrices are given, we can predict a model featuring the CZESs in a arbitrary $d$D system.  Remarkably, when the bulk and boundaries are gapped of the system, we will obtain a $d$th-order topological insulator.


\section{The details and comparisons of different
  cases in  2D system}
When $d=2$ for the 2D Hamiltonian in Eq.~\eqref{ad}, the Eq.~\ref{case} transform into the form
\beqn
\label{cases-2}
&&(\text{i}): \{\Gamma_{ax}^{4},\Gamma_{ay,by}^{4}\}=0, \{\Gamma_{bx}^{4},\Gamma_{ay,by}^{4}\}=0,\nonumber\\
&&(\text{ii}): [\Gamma_{ax}^{4},\Gamma_{ay,by}^{4}]=0, [\Gamma_{bx}^{4},\Gamma_{ay,by}^{4}]=0,\nonumber\\
&&(\text{iii}): [\Gamma_{ax}^{4},\Gamma_{ay,by}^{4}]=0, \{\Gamma_{bx}^{4},\Gamma_{ay,by}^{4}\}=0,\nonumber\\
&&(\text{iv}): [\Gamma_{ax}^{4},\Gamma_{ay}^{4}]=0,  \{\Gamma_{ay}^{4},\Gamma_{by}^{4}\}=0,\nonumber\\
&&\quad\quad \{\Gamma_{ax}^{4},\Gamma_{by}^{4}\}=0, [\Gamma_{ay}^{4},\Gamma_{by}^{4}]=0.
\label{2DC}
\eeqn
 Without loss of generality, we choose $\{\Gamma_x^a, \Gamma_x^b, C_x\}=\{\Gamma_1^{4},\Gamma_2^{4},\Gamma_{21}^{4}\}$. For the different commutation relations in Eq.~\ref{2DC}, we can list all possible choices of $\{\Gamma^a_y, \Gamma^b_y,C_y\}$ as
\beqn
&&(\text{i}): \{\Gamma_{\alpha}^4,\Gamma_{\beta}^4,\Gamma_{\beta\alpha}^4\};\nonumber \\
&&(\text{ii}):\{\Gamma_{\alpha\beta}^4,\Gamma_{\beta\gamma}^4,\Gamma_{\alpha\gamma}^4\};\nonumber \\
&&(\text{iii}):\{ \Gamma_{2\alpha}^4,\Gamma_{2\beta}^4,\Gamma_{\beta\alpha}^4 \}; \nonumber\\
&&(\text{iv}):\{\Gamma_{2\alpha}^4,-\Gamma_{1\alpha}^4,\Gamma_{21}^4\},\{\Gamma_{1}^4,\Gamma_{2}^4,\Gamma_{21}^4\},\nonumber\\
&&\quad\quad \{\Gamma_{2\alpha}^4,\Gamma_{2}^4,-\Gamma_{\alpha}^4\},\{\Gamma_{1}^4,\Gamma_{1\alpha}^4,\Gamma_{\alpha}^4\};
\eeqn
with $\alpha \neq \beta \neq \gamma \in (3,4,5)$. Thus,  the situations   $C_x\neq C_y$ and $C_x = C_y$ classified in the main text correspond to the cases (i-iv) and (iv), respectively. 

\begin{figure}
\centering
\includegraphics[width=6in]{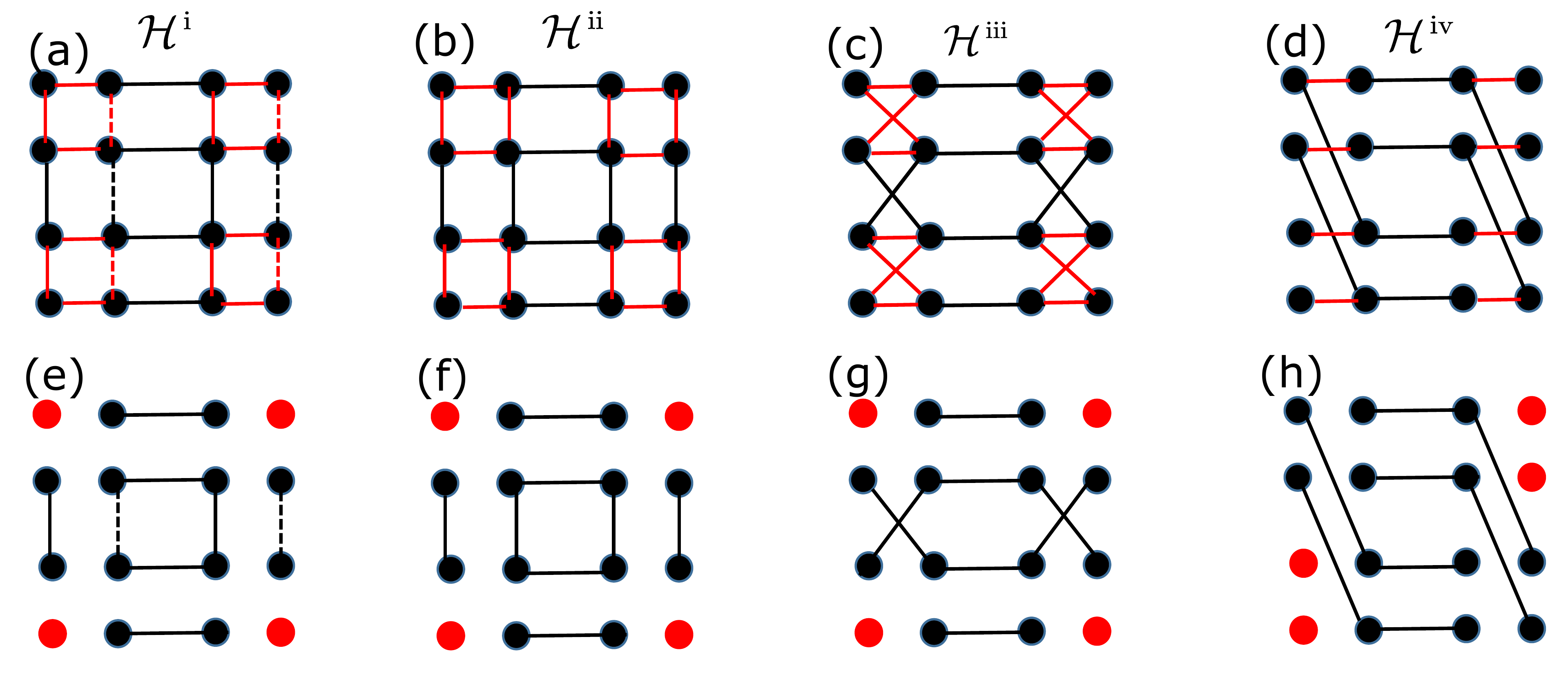}
\caption{ (a)-(d) The schematic diagram of the lattice hoppings for Hamiltonians $\mathcal{H}^{\text{i-iv}}$. The red and black bonds represent the intracellular and intercellular hoppings, respectively. (e)-(h) Schematic of the lattice hoppings in the limit case $t_x=t_y=0$ for Hamiltonians $\mathcal{H}^{\text{i-iv}}$. The existence of the CZESs corresponds to the presence of the isolated atoms at the corners in this limit case. For (e)-(g), the existence of isolated edge atoms coupled in a dimerized way corresponds to the existence of edge states described by the SSH model.}
\label{lat}
\end{figure}

\subsection{The 2D  models predicted by the commutation relations (i-iii) }
Making a classification for 2D Hamiltonian $H(\bm k)(d=2)$ in Eq.\ref{ad}, we can obtain four types commutation relations (i-iv) between the Gamma matrices, as discussed in the main text.  For the case (i-iii), without loss of generality, we consider their representations as
\beqn
&\mathcal{H}^{g}(\bm k)=h_x^{g}(k_x)+h_y^{g}(k_y),\nonumber\\
&h_x^{\text{i}}=M_x(k_x)\tau_x\sigma_0+\lambda_x\sin k_x\tau_y\sigma_0,h_y^{\text{i}}=M_y(k_y)\tau_z\sigma_x+\lambda_y\sin k_y\tau_z\sigma_y, \nonumber\\
& h_x^{\text{ii}}=M_x(k_x)\tau_x\sigma_0+\lambda_x\sin k_x\tau_y\sigma_0,h_y^{\text{ii}}=M_y(k_y)\tau_z\sigma_x+\lambda_y\sin k_y\tau_z\sigma_y,\nonumber\\
&h_x^{\text{iii}}=M_x(k_x)\tau_x\sigma_0+\lambda_x\sin k_x\tau_y\sigma_0,h_y^{\text{iii}}=M_y(k_y)\tau_x\sigma_x+\lambda_y\sin k_y\tau_x\sigma_y,
\eeqn
with $\tau,\sigma$ two sets Pauli matrices and index $g=\text{i,ii,iii}$. It is noted that Hamiltonians $\mathcal{H}^{\text{i}}(\bm k)$ and $\mathcal{H}^{\text{ii}}(\bm k)$ have completely identical topology as the BBH and 2D SSH models and $\mathcal{H}^{\text{iii}}$ denotes the crossed 2D SSH model. For Hamiltonians $\mathcal{H}^{\text{i,ii,iii}}(\bm k)$, we have the chiral symmetries
\beqn
&C_x^{\text{i}}=-\tau_z\sigma_0,C_y^{\text{i}}=-\tau_0\sigma_z,[C_x^{\text{i}},C_y^{\text{i}}]=0,\mathcal{C}^{\text{i}}=C_x^{\text{i}}C_y^{\text{i}}=-\tau_z\sigma_z,\nonumber\\
&C_x^{\text{ii}}=-\tau_z\sigma_0,C_y^{\text{ii}}=-\tau_0\sigma_z,[C_x^{\text{ii}},C_y^{\text{ii}}]=0,\mathcal{C}^{\text{ii}}=C_x^{\text{ii}}C_y^{\text{ii}}=\tau_z\sigma_z,\nonumber\\
&C_x^{\text{iii}}=-\tau_z\sigma_0,C_y^{\text{iii}}=-\tau_0\sigma_z,[C_x^{\text{iii}},C_y^{\text{iii}}]=0,\mathcal{C}^{\text{iii}}=C_x^{\text{iii}}=-\tau_z\sigma_0.
\eeqn
where $[C_s^{g},h_s^{g}]_{+}=0, [\mathcal{C}^{g},\mathcal{H}^{g}]_{+}=0$, with indexs $s=x,y$, $g=\text{i,ii,iii}$. The hopping of these lattice models are schematically shown in Figs.~\ref{lat}(a)-(c).  In Figs.~\ref{lat}(e)-(g), the presence of isolated atoms at the corners in the limit case $t_{x,y}=0$ correspond to the existence of the CZESs.


Remarkably, the band structures of Hamiltonians $\mathcal{H}^{\text{i,ii,iii}}(\bm k)$ can be clearly revealed by  diagonalizing them in the $\sigma$ space as
\beqn
&\mathcal{H}^{\text{i}}(\bm k)=M_x(k_x)\tau_x\sigma_0+\lambda_x\sin k_x\tau_y\sigma_0+E_y\tau_z\sigma_{\varphi}\nonumber,\\
&\mathcal{H}^{\text{ii}}(\bm k)=M_x(k_x)\tau_x\sigma_0+\lambda_x\sin k_x\tau_y\sigma_0+E_y\tau_0\sigma_{\varphi}\nonumber,\\
&\mathcal{H}^{\text{iii}}(\bm k)=M_x(k_x)\tau_x\sigma_0+\lambda_x\sin k_x\tau_y\sigma_0+E_y\tau_x\sigma_{\varphi},
\eeqn
where we have defined $E_y=\sqrt{M_y^2+(\lambda_y\sin k_y)^2}$ and $\sigma_\varphi=\cos\varphi\sigma_x+\sin\varphi\sigma_y$, with $\tan\varphi=\lambda_y\sin k_y/M_y$. Thus, in the eigenbasis of $\sigma_{\varphi}$, $\mathcal{H}^{\text{i,ii,iii}}$ are block-diagonal and two  blocks Hamiltonians are
\beqn
&h_{\pm}^{\text{i}}(\bm k)=M_x(k_x)\tau_x+\lambda_x\sin k_x\tau_y\pm E_y\tau_z\nonumber,\\
&h_{\pm}^{\text{ii}}(\bm k)=M_x(k_x)\tau_x+\lambda_x\sin k_x\tau_y\pm E_y\tau_0\nonumber,\\
&h_{\pm}^{\text{iii}}(\bm k)=(M_x(k_x)\pm E_y)\tau_x+\lambda_x\sin k_x\tau_y,
\label{bd}
\eeqn
with $\pm$ the eigenvalues of $\sigma_{\varphi}$. With given $k_y$, $h_{\pm}^{\text{i}}, h_{\pm}^{\text{ii}}(\bm k),h_{\pm}^{\text{iii}}(\bm k)$ can be viewed as the SSH model along $k_x$, with additional chiral symmetry breaking term $\pm E_y\tau_z$, modulated chemical potential term $\pm E_y\tau_0$, modulated intra-cell hopping term $\pm E_y\tau_x$, respectively.
According to the Eq.~\eqref{bd}, we know that the bulk energy spectrums of Hamiltonians $\mathcal{H}^{\text{i-iii}}(\bm k)$ can be written as
\beqn
&E^{\text{i}}(\bm k)=\pm\sqrt{E_x^2+E_y^2}, E^{\text{ii}}(\bm k)=\pm E_x\pm E_y,\nonumber\\
&E^{\text{iii}}(\bm k)=\pm\sqrt{(M_x\pm E_y)^2+(\lambda_x\sin k_x)^2},
\eeqn
with $E_{x}=\sqrt{M_{x}^2+(\lambda_{x}\sin k_{x})^2}, E_{y}=\sqrt{M_{y}^2+(\lambda_{y}\sin k_{y})^2}$. Thus, for $\mathcal{H}^{\text{i}}(\bm k)$, as long as $E_x\neq 0$ or $E_y\neq 0$, the bulk is full gapped. For $\mathcal{H}^{\text{ii}}(\bm k)$, when $|E_x|_{\text{min}}<|E_y|_{\text{max}}$ or $|E_y|_{\text{min}}<|E_x|_{\text{max}}$, the bulk is fully gapped. Otherwise, the bulk is gapless and behave as a metal. For $\mathcal{H}^{\text{iii}}(\bm k)$, when $||t_x|- |E_y||_{\text{min}}>|\lambda_x|$, the bulk is fully gapped and behave as a trivial insulator or a 2th-order TI when $|t_{x,y}|<|\lambda_{x,y}|$. When $|t_x|+|E_y|_{\text{max}}<|\lambda_x|$, or $||t_x|- |E_y||_{\text{max}}<|\lambda_x|$ and $|t_x|+ |E_y|_{\text{min}}>|\lambda_x|$, the bulk is fully gapped and behave as a weak topological insulator characterized by the edge flat bands, as shown in Figs.~\ref{wti}(a)(b). Otherwise, the bulk is gapless and behave as mirror symmetry protected Weyl semimetal characterized by edge flat bands, as shown in Figs.~\ref{wti}(c)-(f).

 When the bulk is fully gapped, the occupied states for Hamiltonians $\mathcal{H}^{\text{i-iii}}(\bm k)$ can be written as
\beqn
&|\Psi_1^{\text{i}}\rangle=(\sin\theta/2,-\cos\theta/2e^{i\phi})^{{T}}\otimes(1,e^{i\varphi})^{{T}}/\sqrt{2},\nonumber\\
&|\Psi_2^{\text{i}}\rangle=(\cos\theta/2,-\sin\theta/2e^{i\phi})^{{T}}\otimes(1,-e^{i\varphi})^{{T}}/\sqrt{2}\nonumber,\\
&|\Psi_1^{\text{ii}}\rangle=(1,e^{i\phi})^{{T}}\otimes(1,-e^{i\varphi})^{{T}}/2,|\Psi_2^{\text{ii}}\rangle=(1,-e^{i\phi})^{\text{T}}\otimes(1,-e^{i\varphi})^{{T}}/2,\nonumber\\
&|\Psi_1^{\text{iii}}\rangle=(1,-e^{i\beta_1})^{{T}}\otimes(1,e^{i\varphi})^{{T}}/2,|\Psi_2^{\text{iii}}\rangle=(1,-e^{i\beta_2})^{T}\otimes(1,-e^{i\varphi})^{{T}}/2,
\label{bwf}
\eeqn
with $\cos\theta=E_y/|E^{\text{i}}|$, $\tan\phi=\lambda_x\sin k_x/M_x$, $\tan\beta_1=\lambda_x\sin k_x/(M_x+E_y)$, $\tan\beta_2=\lambda_x\sin k_x/(M_x-E_y)$. Here, for the definition of the occupied states, we consider the parameters region $|E_x|_{\text{min}}<|E_y|_{\text{max}}$ for $\mathcal{H}^{\text{ii}}(\bm k)$.

\begin{figure}
\centering
\includegraphics[width=6in]{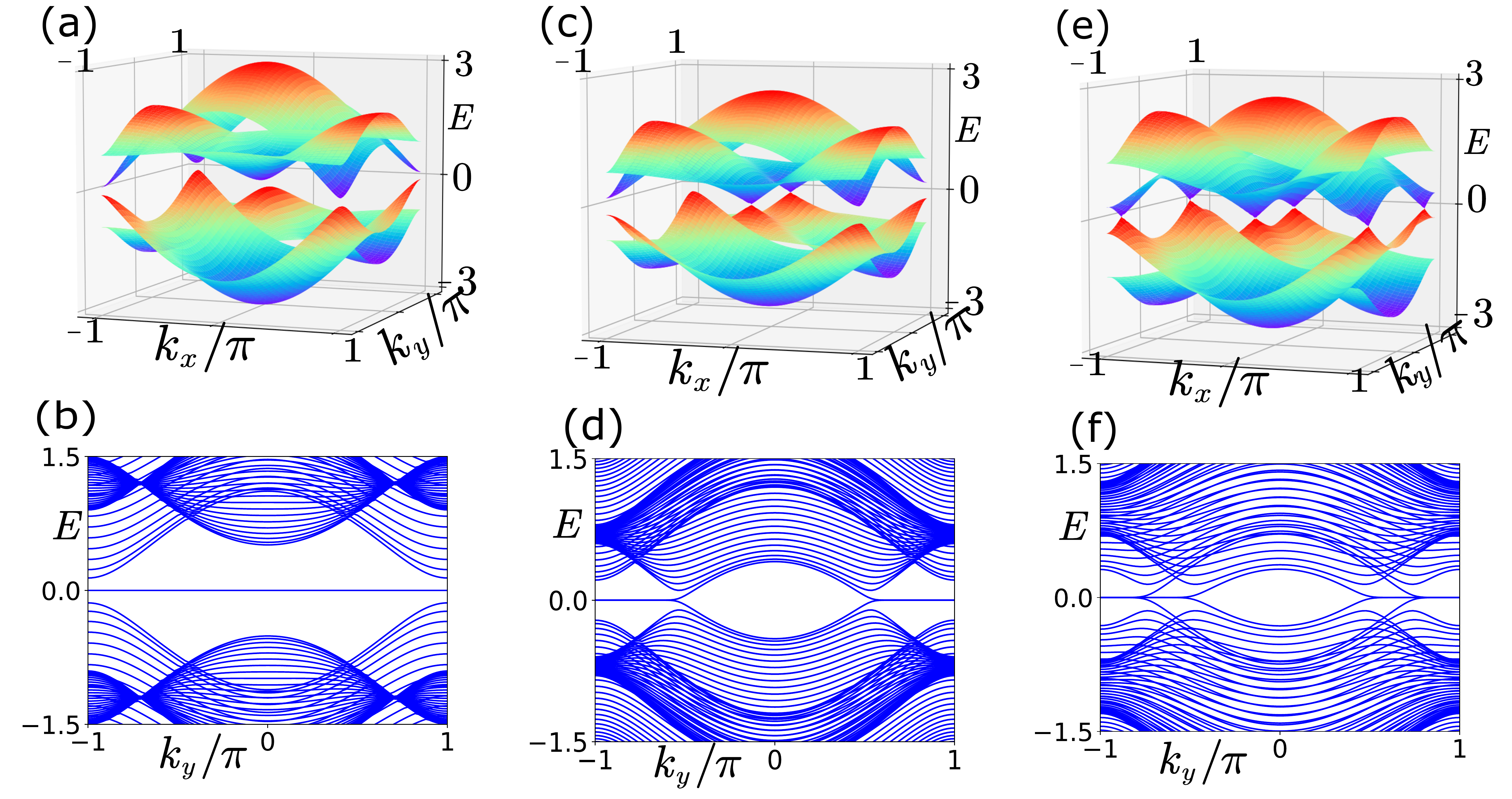}
\caption{ (a)(b) The bulk energy spectrum and edge flat bands of the weak TI phase are plotted. We take  the parameters as $\lambda_x=1,t_x=0.6,\lambda_y=1,t_y=0.5$. (c)(d) The bulk energy spectrum and edge flat bands of the Weyl semimetal with two Weyl points are plotted. We take  the parameters as $\lambda_x=1,t_x=0.6,\lambda_y=0.7,t_y=0.4$. (e)(f) The bulk energy spectrum and edge flat bands of the Weyl semimetal with two Weyl points are plotted. We take the parameters as $\lambda_x=1,t_x=0.6,\lambda_y=1,t_y=0.2$.}
\label{wti}
\end{figure}

\subsection{The comparisons of the topological characterizations}

In the following, we compare the different topological characterizations, including nested Wilson loop, polarization, quadrupole moment, and edge winding number, for the second-order TI phase in cases (i-iii). 
It is known that the CZESs in the BBH model can be characterized by the nested Wilson loop topological invariants, which reflect the topology of the gapped Wannier band. From the bulk wave function in Eq.~\eqref{bwf}, the Wannier bands $\nu(k_y)$, the momentum dependent Berry phase of the occupied states, can be calculated as
\beqn
\nu_{n}^{g}(k_y)=\int_{-\pi}^{\pi}A_n^{g}(\bm k)dk_x=-i\int_{-\pi}^{\pi}\langle\Psi_n^{g}(\bm k)|\partial_{k_x}|\Psi_n^g(\bm k)\rangle dk_x,
\eeqn
with the occupied states index $n=1,2$, $A_n^{g}(\bm k)$ the Berry connection. According to the Eq.~\eqref{bwf}, we have
\beqn
&A_1^{\text{i}}=\cos^2\theta/2\partial_{k_x}\phi,A_2^{\text{i}}=\sin^2\theta/2\partial_{k_x}\phi,\nonumber\\
&A_1^{\text{ii}}=A_2^{\text{ii}}=\partial_{k_x}\phi/2,A_1^{\text{iii}}=\partial_{k_x}\beta_1/2,A_2^{\text{iii}}=\partial_{k_x}\beta_2/2.
\eeqn
After the integration for the Berry connection,  the Wannier bands $\nu^{\text{i}}_1(k_y)=2\pi\cos^2\theta/2, \nu^{\text{i}}_2(k_y)=2\pi\sin^2\theta/2$. Thus, the Wannier bands $\nu_{1,2}^{\text{i}}$ for $\mathcal{H}^{\text{i}}(\bm k)$ are gapped when $E_y\neq 0$. Otherwise when $E_y= 0$, the chiral symmetry for $h_{\pm}^{\text{i}}$ restores and the Wannier bands are gapless at $k_y=0/\pi$, namely $\nu_{1}^{\text{i}}(k_y=0/\pi)=\nu_{2}^{\text{i}}(k_y=0/\pi)$. Thus, the Wannier band $\nu_{n}^{\text{i}}(k_y)$ has the same topological phase transition condition as $h_y$. Analogously, Wannier band $\nu_n^{\text{i}}(k_x)$ has the same topological phase transition condition  as $h_x$. Thus, the CZESs existing condition $\nu_{x,y}=1$ can be extracted from the Wannier band topology through the nested Wilson loop topological invariants. Nevertheless, because $\phi, \beta_1,\beta_2$ are the periodic function of $k_x$, the Berry phases $\nu_{n}^{\text{ii,iii}}(k_y)$ for the occupied states of $\mathcal{H}^{\text{ii,iii}}(\bm k)$ are always quantized to 0 or $\pi$, which indicates that the Wannier bands are gapless for these two cases. Then the nested Wilson loop method fails to characterize the CZESs in cases (ii,iii). Thus, the nested Wilson loop characterizations for the CZEs are only valid for case (i).

\begin{figure}
\centering
\includegraphics[width=6in]{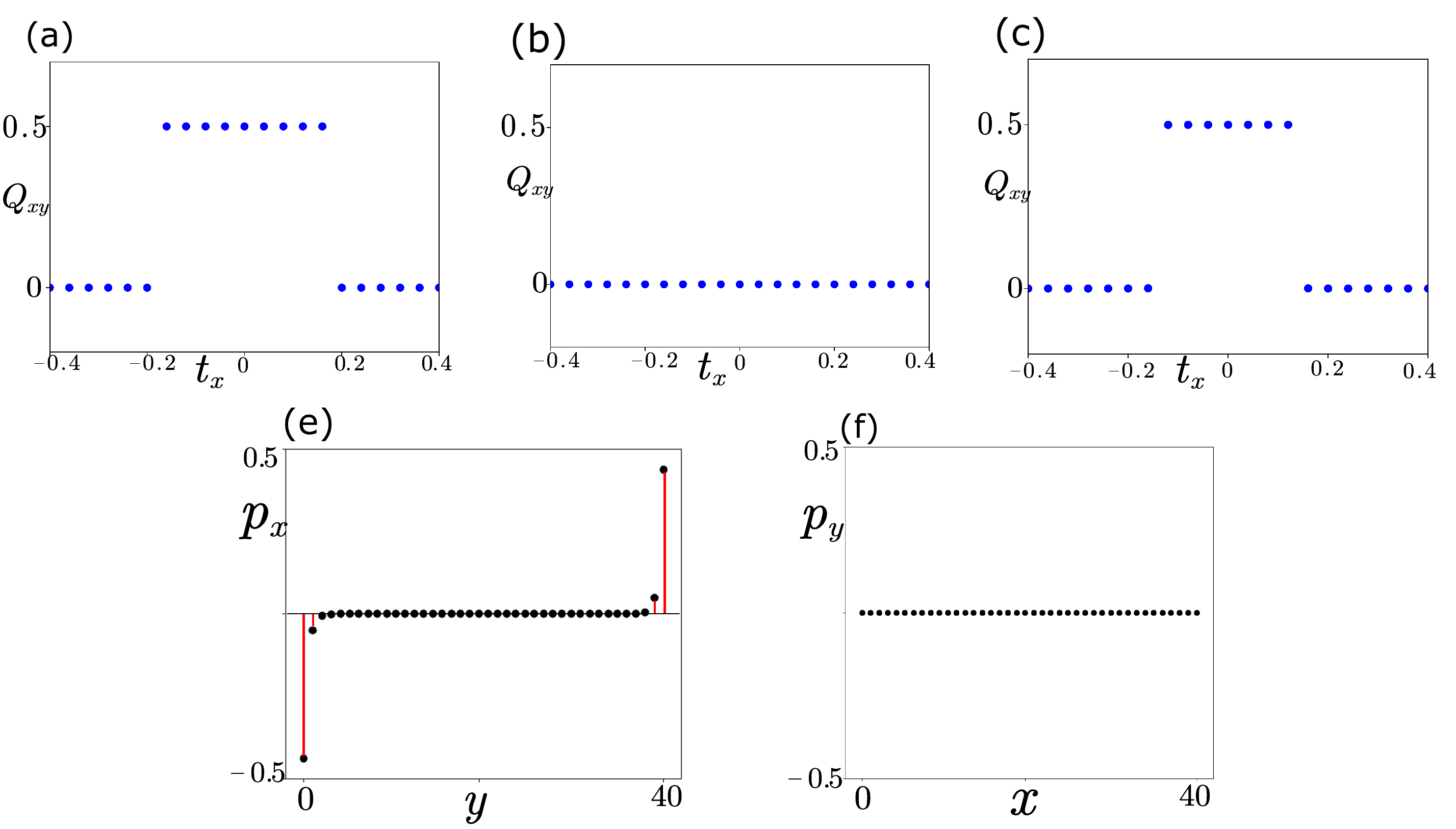}
\caption{ (a)-(c) The quadrupole moment numerical calculations for the fully gapped phases in cases (i-iii). Common parameters are taken as $\lambda_x=0.2,\lambda_y=1,t_y=0.3$. We perform the numerical calculation with the size $41\times 41$. It is noted that the samll deviation  from the exact phase transition point $|t_x|=0.2$ in (c) is resulted by the size effect. (e)(f) The numerical calculations of edge polarizations $(p_x^{\text{edge}},p_y^{\text{edge}})$ for the case (iii). We take the parameters as $t_x=0.1,\lambda_x=0.2,t_y=0.3,\lambda_y=1$.}
\label{QXY}
\end{figure}


For the 2D SSH model, it has been shown that the CZEs can be characterized by the bulk polarization. For example, the polarization along $x$ direction can be written as
\beqn
P_x^{g}&=&\frac{i}{4\pi^2}\sum_n\int dk_ydk_x\text{Tr} [\langle\Psi_n^{g}(\bm k)|\partial_{k_x}|\Psi_n^{g}(\bm k)\rangle\nonumber\\
&=&\frac{1}{4\pi^2}\int dk_y(\nu_{1}^{g}(k_y)+\nu_{2}^{g}(k_y)).
\eeqn
For model $\mathcal{H}^{\text{i}}(\bm k)$, because the chiral symmetry breaking term $E_y\tau_z$ is opposite for $h_{+}^{^{\text{i}}}$ and $h_{-}^{^{\text{i}}}$, we have $\nu_{1}^{\text{i}}(k_y)+\nu_{2}^{\text{i}}(k_y)=2\pi$, giving rise to trivial polarization. Thus, the polarization characterization for the CZEs is invalid for this case.
For model $\mathcal{H}^{\text{ii}}(\bm k)$, $\nu_{1,2}^{\text{ii}}(k_y)$ are both quantized to $\pi$ and $0$ when $\nu_x=1$ and $\nu_x=0$, respectively. The former case leads to nontrivial polarization for each band.
Similarly, $\nu_{1,2}^{\text{ii}}(k_x)$ are both quantized to $\pi$ and $0$ when $\nu_y=1$ and $\nu_y=0$, respectively. Thus the CZEs can be characterized by the polarization of each band for this case. For the second-order TI phase in model $\mathcal{H}^{\text{iii}}(\bm k)$, the
Berry phase $\nu_{1,2}^{\text{iii}}(k_y)$ are both quantized to the value 0, leading to trivial polarization. Thus, the polarization topological invariant also can not characterize the existence of the CZEs for this case.
\begin{figure}
\centering
\includegraphics[width=6in]{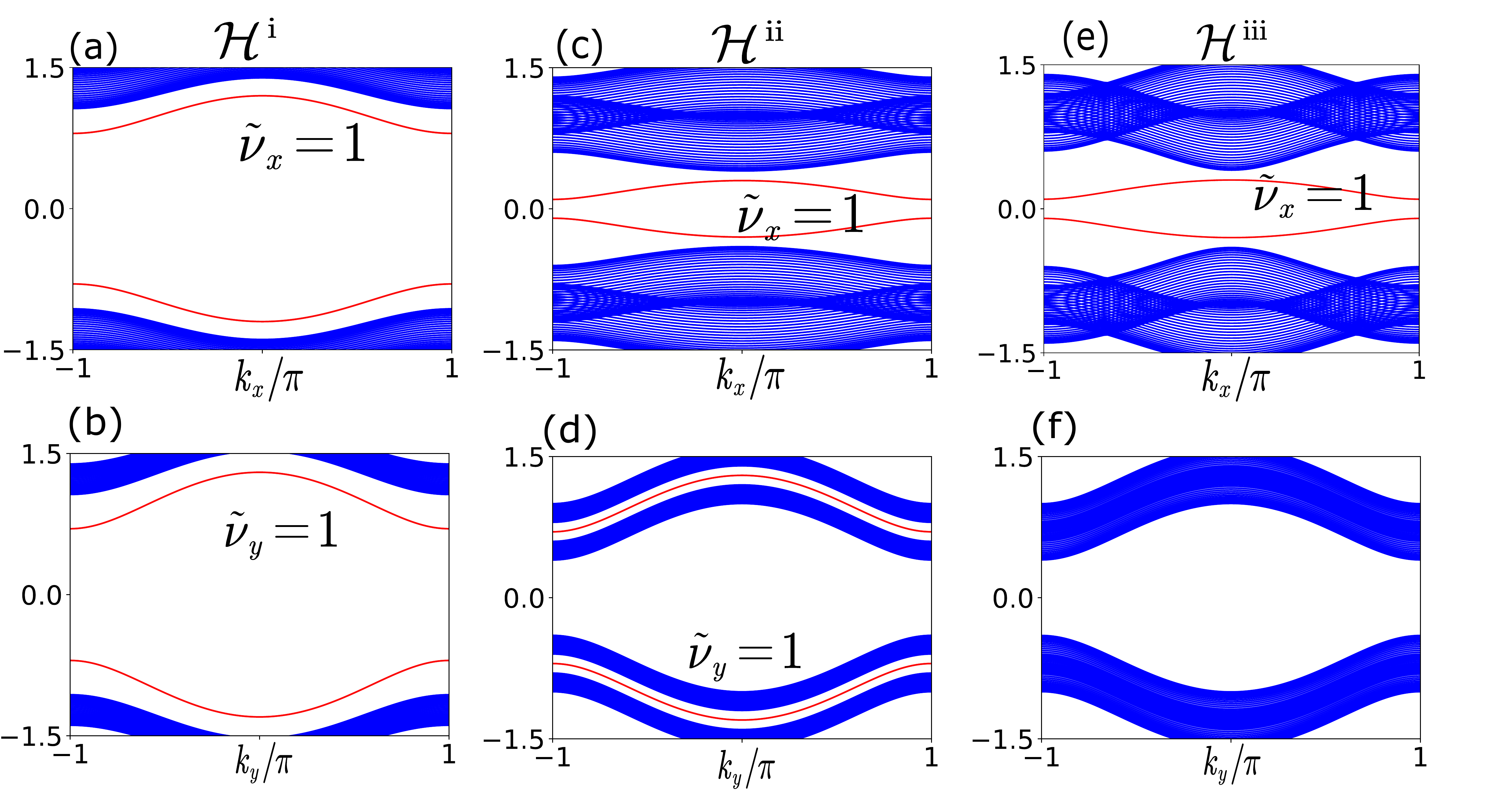}
\caption{ (a)(b) The energy spectrum for $\mathcal{H}^{\text{i}}(\bm k)$ with ribbon geometry along $k_x$ and $k_y$ directions, respectively. We take model parameters as $\lambda_x=\lambda_y=1,t_x=0.2,t_y=0.3$. (c)(d) The energy spectrum for $\mathcal{H}^{\text{ii}}(\bm k)$ with ribbon geometry along $k_x$ and $k_y$ directions, respectively. We take  model parameters as $\lambda_x=0.2,\lambda_y=1,t_x=0.1,t_y=0.3$. (e)(f) The energy spectrum for $\mathcal{H}^{\text{iii}}(\bm k)$ with ribbon geometry along $k_x$ and $k_y$ directions, respectively. We take  model parameters as $\lambda_x=0.2,\lambda_y=1,t_x=0.1,t_y=0.3$. }
\label{edge}
\end{figure}

It is known that quadrupole moment as a higher-order topological invariant can characterize the existence CZEs in the BBH model~\cite{Kang2019,Wheeler2019}. The quadrupole moment can be calculated in real space and it is given by
\beqn
Q_{xy}=[\frac{1}{2\pi}\text{Imlog}[\text{det}(U^{\dagger}\hat{Q}U)]-q_{xy}]\text{mod}1
\eeqn
where the matrix $U$ is constructed by column-wise packing of the occupied eigenstates under the periodic boundary conditions, $\hat{Q}=e^{2\pi i\hat{x}\hat{y}/L_xL_y}$ and $\hat{x},\hat{y}$ are the position operators,
$q_{x y}=\frac{1}{2} \sum_{j=1}^{n} x_{j} y_{j} /\left(L_{x} L_{y}\right)$ is the
contribution from the background positive charge distribution, with $n$ the dimension of the bulk Hamiltonain.
Our numerical calculations show that the quadrupole moment topological invariant $Q_{xy}$ can characterize the CZEs for Hamiltonian $\mathcal{H}^{\text{iii}}(\bm k)$, but can not characterize the CZEs for Hamiltonian $\mathcal{H}^{\text{ii}}(\bm k)$. The numerical results are shown in Figs.~\ref{QXY}(b)(c). Moreover distinct form the BBH model where $p_x^{\text{edge}}=p_y^{\text{edge}}=0.5$, the edge polarizations  $p_x^{\text{edge}}=0.5,p_y^{\text{edge}}=0.$, as shown in Figs.\ref{QXY}(e)(f) in model $\mathcal{H}^{\text{iii}}(\bm k)$. Thus, the second-order TI phase in $\mathcal{H}^{\text{iii}}(\bm k)$ is a phase of type-II quadrupole TI.


In the main text, we have shown that the edge winding number $\tilde{\nu}_x=1$ can completely reflect the CZEs existing condition $\nu_{x,y}=1$. For the BBH and 2D SSH models, there are edge states along $k_x$ and $k_y$ directions, as shown in Figs.\ref{edge}(a)(b) and (c)(d), respectively. Generally, the wave function of the edge states along $k_x$ and $k_y$ can be written as
\beqn
&|\Psi^{g}(k_x,r_y)\rangle_{z_y}=f_{z_y}^{g}(r_y)P_{z_y}^{g}|\psi^{g}(k_x)\rangle,h_x^{g}(k_x)|\psi^{g}(k_x)\rangle=E_x(k_x)|\psi^{g}(k_x)\rangle,\nonumber\\
&|\Psi^{g}(r_x,k_y)\rangle_{z_x}=f_{z_x}^{g}(r_x)P_{z_x}^{g}|\psi^{g}(k_y)\rangle,h_y^{g}(k_y)|\psi^{g}(k_y)\rangle=E_y(k_y)|\psi^{g}(k_y)\rangle,
\eeqn
with the projection operators $P_{z_x}^{g}=(1+z_xC_x^{g})/2$ and $P_{z_y}^{g}=(1+z_yC_y^{g})/2$, g=i,ii. It can be readily verified that
\beqn
\mathcal{H}(k_x,r_y)|\Psi^{g}(k_x,r_y)\rangle_{z_y}=E_x(k_x)|\Psi^{g}(k_x,r_y)\rangle_{z_y},\nonumber\\
\mathcal{H}(r_x,k_y)|\Psi^{g}(r_x,k_y)\rangle_{z_x}=E_y(k_y)|\Psi^{g}(r_x,k_y)\rangle_{z_x},
\eeqn
which means that the edge states $|\Psi^{g}(k_x,r_y)\rangle_{z_y}$ and $|\Psi^{g}(r_x,k_y)\rangle_{z_x}$ have the same energy spectrum as $h_x$ and $h_y$, respectively. Notably, the existence of the edge states along $k_x$ and $k_y$ requires $[C_y^{g},h_x^{g}]=0$ and $[C_x^{g},h_y^{g}]=0$, respectively. Otherwise, $|\Psi^{g}(k_x,r_y)\rangle_{z_y}$ or $ |\Psi^{g}(k_y,r_x)\rangle_{z_x}$ is a null vector after the projection. It can be readily verified that $[C_x^{\text{i,ii}},h_y]=0,[C_y^{\text{i,ii}},h_x]=0$ for both cases (i) and (ii), and $\{C_x^{\text{iii}},h_y\}=0,[C_y^{\text{iii}},h_x]=0$ for case (iii). Thus, there are both edge states along $k_x$ and $k_y$ for cases (i-ii), but there are edge states only along $k_x$ for case (iii), as shown in Figs.~\ref{edge}(e)(f). Correspondingly, the edge Hamiltonian describing these edge states can be obtained by projecting $h_x$ or $h_y$ into the subspace defined by $P_{z_y}$ or $P_{z_x}$. Then we will find that these edge states are described by a SSH model, which consists with the existence of edge isolated atoms coupled in a dimerized way in the limit case $t_x=t_y=0$, as shown in Figs.~\ref{lat}(e)-(g). Remarkably, the existence of edge states along $k_x$ and $k_y$ require that $h_y$ and $h_x$ are topologically nontrivial, respectively. Thus, the edge winding number $\tilde{\nu}_x=1$ or $\tilde{\nu}_y=1$ of the 1D edge states can completely reflect the CZEs existing conditions $\nu_x=\nu_y=1$. As a result, the edge winding number can completely characterize the existence of the CZEs for cases (i-iii), revealing the unified edge-corner correspondence.

\subsection{The  model predicted by the commutation relation (iv)}
In the main text, classifying the 2D system constructed by the SSH model from each direction, we obtain the case (iv) and it can be realized by considering the Hamiltonian
\beqn
&\mathcal{H}^{\text{iv}}(\bm k)=M_x(k_x)\Gamma_{x}^{a}+\lambda_x\sin k_x\Gamma_{x}^{b}+\lambda_yM_y(k_y)\Gamma_{y}^{a}+\sin k_y\Gamma_{y}^{b},\nonumber\\
&\Gamma_{x}^{a}=\tau_z\sigma_x,\Gamma_{x}^{b}=\tau_z\sigma_y,\Gamma_{y}^{a}=\tau_0\sigma_x,\Gamma_{y}^{b}=\tau_0\sigma_y.
\eeqn
The lattice hopping of $\mathcal{H}^{\text{iv}}(\bm k)$ is shown in Figs.~\ref{lat}(d)(h). For this concrete model, we have chiral symmetries $C_x^{\text{iv}}=C_y^{\text{iv}}=-\sigma_z$ and $\Gamma_{x}^{a}\Gamma_{y}^{a}=\tau_z\sigma_0$, which commutes with Hamiltonian $\mathcal{H}^{\text{iv}}(\bm k)$. Correspondingly, $\mathcal{H}^{\text{iv}}$ is block diagonal in $\tau$ space and each block can be written as $h_{\pm}^{\text{iv}}=(\pm M_x+M_y)\sigma_x+(\pm\lambda_x\sin k_x+\lambda_y\sin k_y)\sigma_y$. Here, $h_{+}$ and $h_{-}$ have the identical physics and we focus on the block $h_{+}^{\text{iv}}$. This two bands model can be separated into two 1D Hamiltonians $h_{s=x,y}^{'}=(t_s^{'}+\lambda_s\cos k_s)\sigma_x+\lambda_s\sin k_s\sigma_y$ with $t_x^{'}+t_y^{'}=t_x+t_y=t$. As long as $h_{x}^{'}$ and $h_y^{'}$ are topologically nontrivial with end zero states, the 2D Hamiltonian $h_{+}$ has CZEs localized at the diagonal corners according to our construction principle. Thus, there are CZEs for $h_{+}$ when $|t|<|\lambda_x|+|\lambda_y|$.

Obviously, $h_{+}^{\text{iv}}$ can be viewed as a modulated SSH along $k_{x/y}$, with $k_{y/x}$ given. The band structures can be clearly revealed by the Berry phase $\nu^{\text{iv}}(k_x)$ or $\nu^{\text{iv}}(k_y)$ of the occupied state of $h_{+}$, with $k_x$ or $k_y$ given. When $\nu^{\text{iv}}(k_x)$ or $\nu^{\text{iv}}(k_y)$ is quantized to $\pi$ over all the range, the bulk behave as a weak TI characterized by edge flat band, corresponding to the condition $|t|+|\lambda_x|<|\lambda_y|$ or $|t|+|\lambda_y|<|\lambda_x|$.
When $\nu^{\text{iv}}(k_y)$ and $\nu^{\text{iv}}(k_x)$ are both quantized to $0$ over all the range, corresponding to the condition $|t|>|\lambda_x|+|\lambda_y|$, the bulk is a normal insulator. When $\nu^{\text{iv}}(k_x)$ or $\nu^{\text{iv}}(k_y)$ is not successive, corresponding to the condition $||t|-|\lambda_x||<|\lambda_y|<|t|+|\lambda_x|$ or $||t|-|\lambda_y||<|\lambda_x|<|t|+|\lambda_y|$, the bulk is a Weyl semimetal characterized by edge flat band. Thus, distinct from the case (iii), the predicted CZEs here always coexist with the edge flat band, which brings the difficulty to identify and characterize the predicted CZEs.

\section{The CZEs in the superconducting system}
In the main text, we consider the 2D electronic system constructed by the combination of the SSH model along different directions. By classifying this system, we obtain four topologically unequivalent models supporting the CZEs. In the following, we show that all these models with different commutation relations between the Gamma matrices can be realized in the superconducting system by allowing additional particle-hole symmetry. Directly, considering in the superconducting Bogoliubov–de Gennes (BdG) basis $\Psi(\bm k)=(c_{\uparrow,\bm k},c_{\downarrow,\bm k},c_{\uparrow,\bm k}^{\dagger},c_{\downarrow,\bm k}^{\dagger} )$, the BDG Hamiltonian can be generically written as
\beqn
H_{\text{BDG}}(\bm k)=M_{\text{e}}(\bm k)(\tau_zs_0+\tau_zs_x+\tau_zs_z+\tau_0s_y)+M_{\text{o}}(\bm k)(\tau_0s_x+\tau_0s_z+\tau_zs_y)\nonumber\\
+\Delta_{\text{e}}(\bm k)(\tau_ys_y+\tau_xs_y)+\Delta_{\text{o}}(\bm k)(\tau_xs_0+\tau_xs_z+\tau_xs_x+\tau_ys_0+\tau_ys_x+\tau_ys_z),
\eeqn
where, $\tau,s$ are Pauli matrices in the particle-hole and spin space, respectively. Here, required by the particle-hole symmetry $\mathcal{P}=\tau_xK$, we have $M_e(\bm k)=M_e(-\bm k), M_{\text{o}}(\bm k)=-M_{\text{o}}(-\bm k), \Delta_{\text{e}}(\bm k)=\Delta_{\text{e}}(-\bm k),\Delta_{\text{o}}(\bm k)=-\Delta_{\text{o}}(-\bm k)$. Obviously, all the 15 traceless Dirac matrices can enter into the BDG Hamiltonian. In the following, we consider the BDG Hamiltonian
\beqn
\mathcal{H}_{\text{BDG}}=M_x(k_x)\Gamma_{x}^{a}+\lambda_x\sin k_x\Gamma_{x}^{b}+\lambda_yM_y(k_y)\Gamma_{y}^{a}+\sin k_y\Gamma_{y}^{b},
\eeqn
with
\beqn
&\Gamma_{x}^{a},\Gamma_{y}^{a}\in\{\tau_zs_0,\tau_zs_x,\tau_zs_z,\tau_0s_y,\tau_ys_y,\tau_xs_y\},\nonumber\\ &\Gamma_{x}^{b},\Gamma_{y}^{b}\in\{\tau_0s_x,\tau_0s_z,\tau_zs_y,\tau_xs_0,\tau_xs_z,\tau_xs_x,\tau_ys_0,\tau_ys_x,\tau_ys_z\}.
\eeqn
Under the CZEs existing condition $[i\Gamma_{x}^{a}\Gamma_{x}^{b},i\Gamma_{y}^{a}\Gamma_{y}^{b}]=0$ for the BDG Hamiltonain $\mathcal{H}_{\text{BDG}}$, we will show that all four types commutation relations between the Dirac matrices can be realized.


For the case (i), we require $ \{\Gamma_{x}^{a},\Gamma_{y}^{a,b}\}=0, \{\Gamma_{x}^{b},\Gamma_{y}^{a,b}\}=0$. This case can be realized by considering the represtations
\beqn
&\mathcal{H}_{\text{BDG1}}^{\text{i}}(\bm k)=M_x(k_x)\tau_zs_x+\lambda_x\sin k_x\tau_ys_x+\lambda_yM_y(k_y)\tau_0s_y+\sin k_y\tau_xs_x\nonumber,\\
&\mathcal{H}_{\text{BDG2}}^{\text{i}}(\bm k)=M_x(k_x)\tau_zs_x+\lambda_x\sin k_x\tau_xs_0+\lambda_yM_y(k_y)\tau_zs_z+\sin k_y\tau_ys_0\nonumber,\\
&\mathcal{H}_{\text{BDG3}}^{\text{i}}(\bm k)=M_x(k_x)\tau_ys_y+\lambda_x\sin k_x\tau_0s_x+\lambda_yM_y(k_y)\tau_xs_y+\sin k_y\tau_zs_y,\nonumber\\
&\mathcal{H}_{\text{BDG4}}^{\text{i}}(\bm k)=M_x(k_x)\tau_zs_0+\lambda_x\sin k_x\tau_xs_z+\lambda_yM_y(k_y)\tau_xs_y+\sin k_y\tau_ys_0\nonumber,\\
&\mathcal{H}_{\text{BDG5}}^{\text{i}}(\bm k)=M_x(k_x)\tau_zs_0+\lambda_x\sin k_x\tau_xs_0+\lambda_yM_y(k_y)\tau_ys_y+\sin k_y\tau_ys_x\nonumber,\\
&\mathcal{H}_{\text{BDG6}}^{\text{i}}(\bm k)=M_x(k_x)\tau_zs_z+\lambda_x\sin k_x\tau_0s_x+\lambda_yM_y(k_y)\tau_0s_y+\sin k_y\tau_xs_z.
\eeqn
The bulk states of these Hamiltonians are fully gapped and they behave as the second-order TSCs, which have completely identical topology property as the BBH model. It is noted that the models $\mathcal{H}_{\text{BDG1}}^{\text{i}}(\bm k)$ and $\mathcal{H}_{\text{BDG2}}^{\text{i}}(\bm k)$ have been studied in references~\cite{Wang2018a,Tiwari2020a}. The realization of the model $\mathcal{H}_{\text{BDG3}}^{\text{i}}(\bm k)$ only need the even parity pairings, breaking or preserving time-reversal symmetry ($\mathcal{T}=is_yK$), which has been considered in reference~\cite{Kheirkhah2020}. Besides the model Hamiltonian $\mathcal{H}_{\text{BDG3}}^{\text{i}}(\bm k)$, the realizations of other model Hamiltonians require the $p$-wave pairings, breaking or preserving time-reversal symmetry.

For the case (ii), we require $ [\Gamma_{x}^{a},\Gamma_{y}^{a,b}]=0, [\Gamma_{x}^{b},\Gamma_{y}^{a,b}]=0$. This case can be realized by considering the representation
\beqn
\mathcal{H}_{\text{BDG1}}^{\text{ii}}(\bm k)=M_1(k_1)\tau_zs_0+\sin k_1\tau_xs_x+M_2(k_2)\tau_zs_z+\sin k_2\tau_0s_x.
\eeqn
The realization of this concrete model requires the $p$-wave pairing for the system.

For the case (iii), we require $ [\Gamma_{x}^{a},\Gamma_{y}^{a,b}]=0, \{\Gamma_{x}^{b},\Gamma_{y}^{a,b}\}=0$. This case can be realized by considering the representation
\beqn
\mathcal{H}_{\text{BDG1}}^{\text{iii}}(\bm k)=M_1(k_1)\tau_zs_x+\sin k_1\tau_ys_0+M_2(k_2)\tau_zs_0+\sin k_2\tau_xs_z\nonumber,\\
\mathcal{H}_{\text{BDG2}}^{\text{iii}}(\bm k)=M_2(k_2)\tau_zs_x+\sin k_2\tau_zs_y+M_1(k_1)\tau_ys_y+\sin k_1\tau_0s_x.
\eeqn
The realization of model Hamiltonian $\mathcal{H}_{\text{BDG1}}^{\text{iii}}(\bm k)$ needs the $p$-wave pairing. The realization of model Hamiltonian $\mathcal{H}_{\text{BDG2}}^{\text{iii}}(\bm k)$ only needs the even parity pairing preserving time-reversal symmetry.

For the case (iv), we require $[\Gamma_x^a,\Gamma_y^{a}]=0,  \{\Gamma^1,\Gamma_y^{b}\}=0, \{\Gamma_x^a,\Gamma_y^{b}\}=0, [\Gamma_x^b,\Gamma_y^{b}]=0$. This case can be realized by considering the representation
\beqn
\mathcal{H}_{\text{BDG}}^{\text{iv}}(\bm k)=M_1(k_1)\tau_zs_0+\sin k_1\tau_xs_z+M_2(k_2)\tau_zs_0+\sin k_2\tau_ys_x.
\eeqn
The realization of model Hamiltonian $\mathcal{H}_{\text{BDG}}^{\text{iv}}(\bm k)$ needs the $p$-wave pairing.

\end{widetext}
\end{appendix}

\begin{thebibliography}{82}
\bibitem [{BHZ()}]{BHZ}%
  \BibitemOpen
  \href@noop {} {}\bibinfo {note} {The well-known BHZ model can be separated
  into the form written as a summation of two SSH models along different
  directions, with $\Gamma_{x}^{a}=\Gamma_{y}^{a}$. However, this system does
  not exist the CZSs in real space.}\BibitemShut {Stop}%
\bibitem [{per()}]{per}%
  \BibitemOpen
  \href@noop {} {}\bibinfo {note} {If the perturation $h_{p}$ preserves the
  chiral symmetry $\mathcal{C}$, with $[h_{p},\mathcal{C}]_{+}=0$. Then the
  coupling between the CZSs with same eigenvalue $z$ can be expressed as
  $\langle\Psi_{z}|h_{p}|\Psi_{z}\rangle=z\langle\Psi_{z}|(\mathcal{C}h_{p}+h_{p}\mathcal{C})/2|\Psi_{z}\rangle=0$.}\BibitemShut
  {Stop}%
\bibitem [{MS()}]{MS}%
  \BibitemOpen
  \href@noop {} {}\bibinfo {note} {Because the two bands which cross to form
  the Wely nodes have opposite eigenvalue of mirror-x symmetry
  $\mathcal{M}_x=\tau_x\sigma_0$, the pertubations presvering the mirror
  symmetry can not remove the Wely nodes.}\BibitemShut {Stop}%
\end{thebibliography}
\end{document}